\documentclass[pra,aps,twocolumn,amssymb,showpacs]{revtex4}
\usepackage{graphicx}

\usepackage{epsfig}
\usepackage{graphicx}
\usepackage{dcolumn}
\usepackage{bm}

\begin{document}

\title{  Charge and orbital order in transition metal oxides
         \footnote{\it Dedicated to the memory of the late Professor Jan Stankowski} }

\author{ Andrzej M. Ole\'s }

\affiliation{Marian Smoluchowski Institute of Physics, Jagellonian
             University, Reymonta 4, PL--30059 Krak\'ow, Poland\\
             Max--Planck--Institut f\"ur Festk\"orperforschung,
             Heisenbergstrasse 1, D--70569 Stuttgart, Germany}

\date{22 December 2009}

\begin{abstract}
A short introduction to the complex phenomena encountered in
transition metal oxides with either charge or orbital or joint
charge-and-orbital order, usually accompanied by magnetic order,
is presented. It is argued that all the types of above ordered
phases in these systems follow from strong Coulomb interactions as
a result of certain compromise between competing instabilities
towards various types of magnetic order and optimize the gain of 
kinetic energy in doped systems. This competition provides a natural
explanation of the stripe order observed in doped cuprates,
nickelates and manganites. In the undoped correlated insulators with
orbital degrees of freedom 
the orbital order stabilizes particular types of anisotropic
magnetic phases, and we contrast the case of decoupled spin and
orbital degrees of freedom in the manganites with entangled
spin-orbital states which decide about certain rather exotic
phenomena observed in the perovskite vanadates at finite
temperature. Examples of successful concepts in the theoretical
approaches to these complex systems are given and some open
problems of current interest are indicated.

{\it Published in: Acta Phys. Polon. A {\bf 118}, 212 (2010).}
\end{abstract}


\maketitle

\section{Degrees of freedom in transition metal oxides }
\label{sec:intro}

The physical properties of transition metal oxides are driven by
strong electron interactions \cite{Ima98}. It is due to strong local
Coulomb interactions that these systems exhibit very interesting and
quite diverse instabilities towards ordered magnetic phases when doping
$x$ or temperature $T$ is varied --- is some cases also with orbital
order. These instabilities are observed, {\it inter alia\/}, in rapid
changes of the transport properties at the metal-insulator phase
transitions, or in the onset of superconductivity.

One of the outstanding problems in modern condensed matter theory
is the description of strongly correlated electrons in various
systems. When local Coulomb interactions are strong, the usual
methods used for calculating the electronic structure fail and
have to be extended by the terms following from local interactions, 
either in the framework of the local density approximation (LDA) with
Coulomb $U$, the so-called LDA+$U$ method \cite{Ani91}, or by the
self-energy within the dynamical mean-field theory (DMFT) \cite{Geo96}, 
in the LDA+DMFT approach \cite{Hel07}. This latter approach makes use 
of the local self-energy which becomes exact in the limit of infinite
spatial dimension $d=\infty$ \cite{Met89}. However, even these
methods cannot overcome certain shortcomings of the effective
one-particle theory which justifies modelling of these complex
systems with Hamiltonians of the Hubbard type, and looking for
solutions with methods of quantum many-body theory. The advantage
of rapid progress in the electronic structure calculations in
recent years is that such models can nowadays use realistic
parameters which follow from the electronic structure calculations
for a given system.

Although the field of strongly correlated electronic systems is
very rich, we shall concentrate here on the phenomena observed in
transition metal oxides. There are two major classes of systems
with either perovskite structure $RM$O$_3$, or the layered structure 
$R_2M$O$_4$, with $R$ standing for a rare-earth ion and $M$ for a 
transition metal ion. In the latter class the subsequent layers of
$M$O$_6$ octahedra are displaced, so the electronic properties are
well described by two-dimensional (2D) models, see Ref. \cite{Ima98}. 
In both above structures electron correlations are strong and
lead to remarkable consequences, with several degrees of freedom
contributing simultaneously to coexisting magnetic, charge and (in
some cases also) orbital order. Examples of these complex
phenomena are high-temperature superconductivity \cite{Ben08}, the
colossal magnetoresistance in the manganites
\cite{Dag01,Feh04,Dag05}, and the Verwey transition in the
magnetite (Fe$_3$O$_4$) \cite{Ver39}. Although the charge order occurs 
typically in doped systems, there are a few systems of formally mixed
valence type, where the electron number per one transition metal
ion is not an integer but local correlations stabilize charge
order, as in the magnetite. The latter problem was
recently addressed and the mechanism of the Verwey transition was
explained as triggered by the electron-phonon coupling enhanced by
local Coulomb correlations \cite{Pie06}, so we shall not discuss
it here but refer an interested reader to another contribution in
the same volume \cite{Pie10}.

The electronic structure of transition metal oxides includes
several bands \cite{Ima98,Sol06}, but the properties of the system
do depend on the states in the vicinity of the Fermi energy. It is
usually sufficient to derive effective $M$--O--$M$ hopping
elements for $\sigma$-bonds $t_\sigma$ and $\pi$-bonds $t_\pi$,
and next use them in the effective model describing only $3d$
electrons \cite{Zaa93}. The respective kinetic energy is described
in a perovskite system by
\begin{equation}
\label{H0}
H_{0}=\sum_{\langle ij\rangle,\alpha\beta\sigma}t_{\sigma,\alpha\beta}
a^\dagger_{i\alpha\sigma}a^{}_{j\beta\sigma}
+t_\pi\sum_{\langle ij\rangle,\mu\sigma}
a^\dagger_{i\mu\sigma}a^{}_{j\mu\sigma}\,.
\end{equation}
Here $\{\alpha,\beta\}=\{x,z\}$ are the indices of $e_g$ orbitals,
\begin{equation}
|x\rangle \equiv (x^2-y^2)/\sqrt{2}, \hskip .7cm
|z\rangle \equiv (3z^2-r^2)/\sqrt{6}\,,
\label{eg}
\end{equation}
and this orbital flavor is in general not conserved along the hopping
processes --- the orbitals may be changed for the hopping along the
bonds in $ab$ planes in the perovskite structure. In contrast, the
$t_{2g}$ orbital flavor,
\begin{equation}
|a\rangle \equiv |yz\rangle, \hskip .7cm
|b\rangle \equiv |zx\rangle, \hskip .7cm
|c\rangle \equiv |xy\rangle.
\label{t2g}
\end{equation}
is conserved for the hopping along the bonds in all three cubic
directions $\gamma=a,b,c$, as indicated by a single diagonal
hopping element $t_\pi$ (for simplicity we assume only nearest
neighbor hopping elements), with $\mu=\{a,b,c\}$ labeling $t_{2g}$
orbitals and referring to the cubic axes perpendicular to the
planes accomodating the respective orbitals. The latter notation is
introduced using an ideal cubic system in which each $t_{2g}$
orbital is perpendicular to a single cubic axis, for instance the
$|xy\rangle$ orbital lies in the $ab$ plane and is perpendicular
to the $c$ axis.

On-site intraorbital Coulomb interactions are described by a single
parameter $U$ (identical for all $3d$ orbitals):
\begin{equation}
\label{H_U}
H_{U}=U\sum_{i\alpha}n_{i\alpha \uparrow}n_{i\alpha\downarrow}\,.
\end{equation}
In the simplest approach, the ratio $U/W$, where $W$ is the
bandwidth for the relevant partly filled band, decides whether
electrons localize and the electronic structure changes to two
Hubbard subbands in a Mott insulator, or the system is metallic,
with rather strongly correlated electrons and possibly heavy
effective masses (this happens for the $f$-electron systems which
are addressed in other contributions in this volume). As the
hopping elements along $\pi$ bonds are significantly lower than the 
ones for $\sigma$ bonds \cite{Zaa93}, the $t_{2g}$ electrons in the 
early transition metal oxides (i.e. in titanium or vanadium oxides) 
are even stronger correlated than $e_g$ electrons in the $R$MnO$_3$ 
or LaNiO$_3$ perovskites. This resembles the situation in molecular
bonds in $sp$ systems, with $\pi$ bonds being always stronger
correlated than $\sigma$ bonds \cite{Ole86}.

A second class of correlated insulators, so-called charge transfer
insulators, arises when the oxyges states are within the gap between
the two Hubbard subbands \cite{Zaa85}. A crucial parameter is the
energy difference between the $d$ and $p$ electron (hole) levels,
$\Delta=\varepsilon_p-\varepsilon_d$ --- here we use the hole notation
relevant for the high-$T_c$ cuprates. When $\Delta>U$ one has a
Mott-Hubbard insulator, but when $\Delta<U$, the insulator is of
charge transfer type.

The electronic structure of the cuprates does not involve orbital
degeneracy as the CuO$_6$ octahedra are elongated and the orbital
degeneracy is removed for a tetragonal distortion. Therefore, s
hole in the $d^9$ configuration
occupies the $|x\rangle\equiv(x^2-y^2)/\sqrt{2}$ orbital at each
Cu$^{2+}$ ion in La$_2$CuO$_4$.
The resulting charge transfer model for the CuO$_2$ planes in the
cuprates may be thus written as follows \cite{Ole87}:
\begin{equation}
H_{dp}= H_0 + H_{\rm int} \,,
\label{eq:ct} \\
\end{equation}
\begin{eqnarray}
H_0&=&\varepsilon_p\sum_{i}n_{pi}
-t_{pd} \sum_{\langle mi\rangle\alpha\sigma}\gamma_{mi}
\Big( d^\dag_{m \alpha \sigma}p_{i \alpha \sigma}^{} +\mbox{H.c.} \Big)
\nonumber \\
&-&t_{pp} \sum_{\langle ij\rangle \alpha  \sigma}\eta_{ij}
\Big( p^\dag_{i \alpha \sigma}p_{j \alpha \sigma}^{} +\mbox{H.c.} \Big),
\\
H_{\rm int}&=&U_d\sum_{m}n_{m\uparrow}n_{m\downarrow}
+U_p\sum_{i}n_{pi\uparrow}n_{pi\downarrow}\,.
\end{eqnarray}
The parameters of the charge transfer model (\ref{eq:ct}) are: the
oxygen energy $\varepsilon_p$ (we assume that the reference $d$ hole
energy $\varepsilon_d=0$), the $d-p$ hybridization $t_{pd}$, and the
Coulomb interaction parameters for $d$ and $p$ orbitals, $U_d$ and
$U_p$; the same parameters describe also other Cu--O systems, as for
instance CuO$_3$ chains in YBa$_2$Cu$_3$O$_{6+x}$ \cite{Ole91}, or
Cu$_2$O$_5$ coupled ladders in Sr$_{14-x}$Ca$_x$Cu$_{24}$O$_{41}$
\cite{Woh07}. Here $n_{pi}=n_{pi\uparrow}+n_{pi\downarrow}$ and
$n_{pi\sigma}=p_{i\sigma}^{\dagger}p_{i\sigma}^{}$ are charge density
operators, $\gamma_{mi}$ and $\eta_{ij}$ are the phase factor for a
pair of orbitals along the considered $d-p$ ($p-p$) bond.
The parameters for the cuprates which follow from the electronic
structure calculations are (in eV) \cite{Gra92}:
$\Delta=3.6$, $t_{pd}=1.3$, $t_{pp}=0.65$, $U_d\simeq 10.5$,
$U_d\simeq 4.0$. Electron correlations are moderate
in spite of the large value of $U_d$ \cite{Ole87}, but they suffice
to localize holes at Cu sites in the undoped system, such as
La$_2$CuO$_4$ or YBa$_2$Cu$_3$O$_6$. Taking the above parameters,
$\Delta\ll U_d$ and these systems are charge transfer insulators, 
in contrast to the perovskite titanates and vanadates, which are 
Mott-Hubbard systems.

It is important to realize that the charge transfer gap $\Delta$ plays
the role of an effective Coulomb parameter $U\equiv \Delta$ 
in the correlated electronic structure of a charge transfer insulator.
When the Cu--O--Cu hopping between two $|x\rangle$
orbitals along a bond in an $ab$ plane is defined as
$t$, this leads to the effective Hubbard model \cite{Hub63},
\begin{equation}
\label{HM}
H=-t\sum_{\langle ij\rangle,\sigma}
(a^\dagger_{i\sigma}a^{}_{j\sigma}+a^\dagger_{j\sigma}a^{}_{i\sigma})
+U\sum_{i\alpha}n_{i\alpha \uparrow}n_{i\alpha\downarrow}\,.
\end{equation}
For the considered case of $|x\rangle$ orbitals the phase factors
on each bond $\langle ij\rangle$ are identical. Note that $t$ may be
deduced from the charge transfer model (\ref{eq:ct}),
$t=t_{pd}^2/\Delta=0.4$ eV, and for the actual ratio $U/t=10$ the
holes are strongly correlated. Hence, the undoped systems
La$_2$CuO$_4$ or YBa$_2$Cu$_3$O$_6$ are antiferromagnetic (AF)
insulators. In general, the derivation of an effective model from
the relevant multiband model is rather tedious --- such a more
complete model includes in addition next nearest (second) neighbor 
and third nearest neighbor hopping elements $\{t',t''\}$ and 
intersite Coulomb interactions \cite{Fei95}.

A broad class of phenomena investigated for strongly correlated
electron systems are the changes of their physical properties in the
vicinity of metal-insulator transitions. As mentioned above, one way
of localizing electrons in a correlated insulator is by changing the
electron interaction parameter $U$ in Eq. (\ref{HM}) (or the charge 
transfer gap $\Delta$). Although this may be easily realized only in 
theory, in certain systems the changes of the electronic parameters are 
sufficient to induce metal-insulator transitions observed in V$_2$O$_3$ 
\cite{v2o3}.
A more common situation, however, is encountered in doped systems,
where the carriers are released at certain doping concentration and
the system becomes metallic. In contrast to the earlier suggestions,
the one-band model is not sufficient to describe the metal-insulator
transition in V$_2$O$_3$ \cite{Hel01}, and doping is not equivalent to
varying external pressure \cite{Rod10}. This and other metal-insulator
transitions in the oxides are controlled by doping. A very well known
example is the colossal magnetoresistance effect in the perovskite 
manganites \cite{Dag01}, another is the superconductivity in doped 
La$_{2-x}$Sr$_x$CuO$_4$ or YBa$_2$Cu$_3$O$_{6+x}$ compounds \cite{Ben08}.
Other examples can be found, for instance, in the excellent review
article by Imada, Fujimori and Tokura \cite{Ima98}.

In this paper we address in particular the phenomena related to magnetic
and orbital order in transition metal oxides which follow from strong
electron correlations. Charge order arises in doped systems, while
the orbital order is common in transition metal oxides with partly filled
degenerate orbitals. We begin in Sec. \ref{sec:cup} with the stripe
phases in the cuprates, where we explain the stabilizing mechanism and
show that the charge modulation is the way to optimize total energy in
doped systems. While the properties of an undoped cuprate are driven by
the AF superexchange, the systems with orbital superexchange 
interactions are more complex as the interactions are intrinsically 
frustrated \cite{Fei97}. These interactions are exemplified by the 
so-called compass model \cite{Kho03}, see Sec. \ref{sec:orb}, and may 
give either highly degenerate ordered ground states, or the disordered 
orbital liquid. Consequences of the orbital superexchange for the 
magnetic order are addressed in Sec. \ref{sec:som}, where we briefly 
summarize the structure of the spin-orbital superexchange \cite{Ole05}, 
and demonstrate that spin and orbital degrees of freedom may be 
separated in the perovskite manganites. In contrast, in the perovskite 
vanadates spin-orbital entanglement plays a dominating role and decides 
about their properties at finite temperature, see Sec. \ref{sec:van}. 
Finally, we give examples of coexisting charge-and-orbital order in 
doped systems in Sec. \ref{sec:co}. A summary and some open problems 
in the field are given in Sec. \ref{sec:summa}. Figures illustrating 
the theoretical concepts reviewed in this article will not be 
reproduced here --- they may be found in the cited literature which 
is far from being complete and was selected on the criterion of 
addressing the most important concepts in this field.

\section{Stripe phases in the cuprates}
\label{sec:cup}

A crucial concept in the physics of the superconducting cuprates is
the Zhang-Rice singlet \cite{Zha88}. It makes an explicit use of the
charge transfer nature of the electronic structure, as a doped hole
occupies not a Cu($d_x$) orbital but a linear combination of
$p_{\sigma}$ orbitals with $x^2-y^2$ symmetry around a hole, which forms
a singlet together with the hole at Cu ion. It is this concept which
provides a justification for using the $t$--$J$ model as the effective
model describing the physical situation in the cuprates, and plays a
prominent role in this class of compounds \cite{Arr09}.

The $t$-$J$ model itself was derived from the Hubbard model in Cracow
more than three decades ago \cite{Cha77}, using the perturbation theory.
A properly chosen canonical transformation leads from the full Hilbert
space to an effective low-energy Hamiltonian acting in the restricted
space, where only spins and holes occur at different sites. It consists
of the kinetic energy $\propto t$ and the superexchange interaction
$\propto J$ between $S=1/2$ spins:
\begin{equation}
\label{tJ}
H_{t-J}\!=\!-t\!\sum_{\langle ij\rangle,\sigma}(
{\tilde a}^\dagger_{i\sigma}{\tilde a}^{}_{j\sigma}
+{\tilde a}^\dagger_{j\sigma}{\tilde a}^{}_{i\sigma})
+J\sum_{\langle ij\rangle}\left({\bf S}_{i}\!\cdot\!{\bf S}_{j}\!+\!
\frac{1}{4}{\tilde n}_i{\tilde n}_j\right)\!,
\end{equation}
with the superexchange interaction,
\begin{equation}
J=\frac{4t^2}{U}\,.
\label{J}
\end{equation}
The operators ${\tilde a}^\dagger_{i\sigma}=a^\dagger_{i\sigma}(1-
n^{}_{i\bar{\sigma}})$ ($\bar{\sigma}=-\sigma$) are projected fermion
operators and act in the restricted space. The above $t$--$J$ model may
also be derived directly from the charge transfer model --- in this
(realistic for the cuprates) case the superexchange includes both the
Anderson and charge transfer excitations \cite{Zaa88}. For the cuprates
one finds $J\simeq 0.13$ eV, which is either deduced from the magnetic
experiments \cite{Bir98}, or derived from the charge transfer
model using its parameters \cite{Gra92}.

The first intriguing question concerning hole doping is whether a
doped hole may propagate coherently in the antiferromagnet. Naively
one might argue that a hole creates defects on its way, so it
would need to make a hopping along a closed loop to annihilate
these defects and to move in the square lattice with a minute
dispersion \cite{Tru88}. Actually, this is the only process by
which a hole may delocalize in the Ising model. The situation is
quite different, however, when a hole is doped into a Heisenberg
antiferromagnet --- in this case the quantum fluctuations of the
AF background may repair the defects created by the hole, and the
hole dispersion occurs on the energy scale of $J$
\cite{Kan89,Mar91}. This concept was confirmed by experiment, and
indeed the hole dispersion on the low energy scale of $J$ was
observed in the cuprates \cite{Dam03}. Detailed comparison between
the experimental data of angle resolved photoemission experiments
and the outcome of the theoretical calculations performed using
the self-consistent Born approximation (SCBA) \cite{Mar91} were
presented by several groups. Here we mention only the experimental
data of high quality obtained for Sr$_2$CuO$_2$Cl$_2$ by Wells
{\it et al.} \cite{Wel95}, which fit very well to the theoretical
curves obtained with finite next neighbor hopping $t'$
\cite{Bal95}. This demonstrates that the right effective model for
the high-$T_c$ cuprates is the $t$-$t'$-$J$ rather than the $t$--$J$
one.

Higher doping of CuO$_2$ planes leads to a gradual weakening of AF
correlations, which however survive even in the overdoped regime at
$x\simeq 0.2$ \cite{Bir98}. There are several possibilities concerning
the phase diagram of doped cuprates \cite{Dag05}, but the commonly
accepted point of view now is that doped holes self-organize in form of
phases with charge modulation \cite{Kiv03}. Such structures with 
coexisting charge and magnetic order, called {\it stripe phases\/}, 
were first discovered in the theory as an instability of doped 
antiferromagnets towards AF domains separated by (usually) nonmagnetic 
domain walls \cite{Zaa89}. Only a few years later their existence in 
the cuprates was confirmed in the neutron experiments of Tranquada 
{\it et al.} \cite{Tra96}. The stripe phases
are characterized by the coexisting charge and magnetic order, with the
charge density varying twice faster than the spin density in the real
space \cite{Ole00b}.

The first question concerning stripe phases is whether they would form
as solitonic defects in the AF structure, i.e. in between different
AF domains, or instead they are of polaronic nature  not disturbing
the AF order. Although naively one could argue that the polaronic
mechanism could give a better kinetic energy, this argument is 
misleading. To see this one can consider a cluster of three sites 
centered at the domain wall, filled by two electrons \cite{Ole00b}. 
Due to strong correlations with $U\gg t$, the particles are confined 
in this cluster, although it is just a part of the AF 2D plane. Taking 
two electrons with either identical spins or with opposite spins, it 
is straightforward to estimate the ground state energy of two possible 
configurations: ($i$) polaronic ($E_P$) and ($ii$) solitonic ($E_S$) 
one. One finds that the solitonic energy is
lower by the superexchange energy $J$ which arises from the three-site
hopping terms in this cluster \cite{Ole00b}, with:
$E_P=-\sqrt{2}t$ and $E_S=-\sqrt{2}t-4t^2/U$. This simple argument
explains the experimental finding that charge walls separate AF domains
with different phase of the order parameter.

Quantitative results for the stripe phases were first obtained using
the Hartree-Fock (HF) approximation \cite{Zaa89,Zaa96}, and then refined
using variational wave functions \cite{Gor99}, within DMFT for the 
stripe phases \cite{Fle00,Fle01}, and slave-boson approach 
\cite{Rac06b}. As usually, the HF serves only as a hint for possible
instabilities, and gives remarkably robust stripe structures 
\cite{Zaa96} with rather large amplitude of the charge density between 
the domain wall and the centers of AF domains, and the filling of half 
of doped hole per one stripe charge unit cell, as observed in 
experiment. These stripes are vertical (or horizontal), meaning that 
domain walls are along (10) [or (01)] direction, and insulating. 
Actually, their stability follows from a small gap which opens in the 
electronic structure. This mechanism is subtle and involves certain 
additional modulation, either spin or charge density wave, along the 
domain walls \cite{Zaa96}, so one has to expect major changes when 
electron correlations are implemented. However, variational calculations 
confirmed this picture to some extent \cite{Gor99}, although the 
question whether the stripes are insulating or not could not be resolved 
(following the HF results, it was believed
for a long time that the stripes are insulating).

Stable stripes were also found using an exact diagonalization method
within the DMFT for the two-dimensional Hubbard model, in the broad 
doping range $0.03<x<0.2$ in La$_{2-x}$Sr$_x$CuO$_4$ \cite{Fle00}. 
These calculations allowed also to reproduce the observed crossover 
from diagonal (11) to vertical (01) site-centered stripes at doping
$x\simeq 0.05$  \cite{Yam98}. In addition, also the doping
dependence of the size of magnetic domains and chemical potential shift
$\Delta\mu\propto -x^2$ were found to be in quantitative agreement
with the experimental results for La$_{2-x}$Sr$_x$CuO$_4$. In this way
the paradigm of insulating stripe phases was abolished --- the chemical
potential was varying with doping within the metallic phase.

The spectral functions obtained within the DMFT \cite{Fle01} show
a coexistence of the incoherent states in the lower Hubbard band
and a coherent quasiparticle (QP) close to the Fermi energy. The
main features of the spectra are: a flat part of the QP band near
the $X=(\pi,0)$ point, and gaps for charge excitations at the
$Y=(0,\pi)$ and $S=(\pi/2,\pi/2)$ points in the low-doping regime
$x<1/8$. These gaps are gradually filled and close under
increasing doping, in agreement with the experimental data for
La$_{2-x}$Sr$_x$CuO$_4$ obtained using angle resolved photoemission
\cite{Ino00}. In the range of low temperatures $T$ the obtained spectra 
have a distinct QP peak at the $X=(\pi,0)$ point, present just below 
the Fermi energy $\mu$, and a charge gap and well defined QP at the 
$S=(\pi/2,\pi/2)$ point \cite{Fle01}. 
These calculations demonstrated the importance of dynamical
correlations which strongly screen the local potentials resulting
from on-site Coulomb interactions and lead thus to drastic changes
in the distribution of spectral weight with respect to the HF
picture. It was also shown that the melting of stripe order is
influenced by the second neighbor hopping element $t'$, which
plays also an important role in explaining the observed difference
in the spectral properties between Bi$_2$Sr$_2$CaCu$_2$O$_{8+y}$
\cite{She95} and La$_{2-x}$Sr$_x$CuO$_4$ \cite{Ino00}. At the same
time, $t'$ can tip the energy balance between the filled diagonal
and half-filled vertical stripes \cite{Rac06b}, which might
explain a change in the spatial orientation of stripes observed in
the high $T_c$ cuprates at the doping $x\simeq 1/16$.

More insight into the charge and magnetization distribution as well as
into the stability of stripe phases could be obtained using a
rotationally invariant version of the slave-boson approach in spin 
space in the 2D Hubbard model \cite{Rac06b}. This approach allowed
one to treat strong electron correlations in the stripe phases with
large unit cells relevant in the low doping regime, and gave results
representative of the thermodynamic limit. It also helped to resolve 
the longstanding controversy concerning the role played by the kinetic 
energy in stripe phases. While the transverse hopping across the
domain walls yields the largest kinetic energy gain in the case of the
insulating stripes with one hole per site, the holes propagating
{\it along\/} the domain walls stabilize the metallic vertical (01)
stripes with one hole per two sites, as found in the cuprates.

Recently observed pattern of unidirectional domains in
high-$T_c$ superconductors \cite{Koh07} motivated also search for
coexisting charge modulation and $d$-wave superconductivity. Indeed,
half-filled charge domains separated by four lattice spacings were
obtained along one of the crystal axes leading to modulated
superconductivity with {\it out-of-phase\/} $d$-wave order parameters
in neighboring domains \cite{Rac07}. Both renormalized mean-field (MF)
theory and variational Monte Carlo calculations yield that the energies 
of modulated and uniform phases are very close to each other,
so modulated phases could easily be stabilized by other effects
going beyond the $t$--$J$ model used in these calculations. Novel doped 
phases with superconductivity coexisting with charge modulation or even 
the valence-bond solid order were also reported recently \cite{Rac08}.

\section{Intrinsic frustration of the orbital superexchange}
\label{sec:orb}

In the undoped transition metal compounds the physics is however
frequently not so simple as in CuO$_2$ planes of high $T_c$ 
superconductors, where the superexchange stabilizes the AF long-range 
order. This happens in particular when the orbital degrees of freedom 
are active (when degenerate orbitals are only partly filled) and 
contribute to the magnetic order \cite{Kug82,Cas78}.
The central property of the orbital degrees of freedom is that they
are intrinsically frustrated \cite{Fei97,Tok00}, so they may lead to
novel (ordered or disordered) phases.

Frustration in magnetic systems may be of geometrical origin if
only nearest neighbor interactions are present, or may arise due
to competing exchange interactions \cite{Diep,Nor09}. For
instance, when one considers FM interactions along every second 
vertical line in the square lattice while all other interactions are
AF --- then this 2D Ising model is exactly solvable and has a lower
critical temperature \cite{Lon80} than the one with isotropic
exchange interactions. Frustration for quantum spins acts to enhance 
the effects of quantum fluctuations, leading to a number of different 
types of magnetically disordered states, among which some of the more 
familiar ones are static and resonating valence--bond
(VB) phases \cite{Nor09}. However, also ordered phases may emerge
in systems with frustrated spin interactions from their disordered
manifolds of states, and their mechanism of stability is nowadays
called ``order--by--disorder'' \cite{Diep}. Numerous materials are
known at present whose physical properties could be understood only 
by employing microscopic models with frustrated spin interactions in
which some of these theoretical concepts are exemplified.

A prototype model to study frustration in pseudospin systems which
mimic the directional orbital superexchange \cite{vdB99} is the
2D compass model \cite{Kho03}
\begin{equation}
{\cal H}_{2D}=\sum_{\langle ij\rangle\parallel a} J_x\tau_{i}^x
\tau_{j}^x +\sum_{\langle ij\rangle\parallel b} J_z\tau_{i}^z
\tau_{j}^z \,. \label{com2D}
\end{equation}
In this model the $\tau^x_i\tau^x_j$ interactions $\propto J_x$
for horizontal bonds $\langle ij\rangle$ (along the $a$ axis)
compete with the $\tau^z_i\tau^z_j$ ones $\propto J_z$ in the
vertical direction (along the $b$ axis). Recently the structure of
eigenstates in this model was investigated by numerical methods
\cite{Mil05}, and it was shown using quantum Monte Carlo that a
phase transition at finite temperature exists in the 2D compass
model \cite{Wen08}, suggesting that this model is indeed in the 2D
Ising universality class. A competition of pseudospin interactions
along different directions results here in intersite correlations
similar to the anisotropic XY model, and in competition between
two types of Ising-like order. This competition culminates in the
highly degenerate ground state at the compass point (i.e., when
all interactions have the same strength) \cite{Mil05}, and
generates there a first order phase transition when the
anisotropic model with $J_z>J_x$ changes into $J_z<J_x$ through
the $J_z=J_x$ transition point \cite{Vid09}. It is interesting to
note that a similar first order quantum phase transition occurs
also in the one-dimensional (1D) compass model \cite{Brz07}, when
both above interactions alternate along the chain ($N'=N/2$ is the
number of unit cells):
\begin{equation}
{\cal H}_{1D}=\sum_{i=1}^{N'}
\left\{ J_x\tau_{2i-1}^x \tau_{2i}^x +
        J_z\tau_{2i}^z   \tau_{2i+1}^z \right\}\,.
\label{com1D}
\end{equation}
This model was solved exactly in the entire range of $\{J_x,J_z\}$
parameters \cite{Brz07} by mapping onto the exactly solvable
quantum Ising model \cite{Perk} in different subspaces. Equal
coupling constants $J_x=J_z=J$ correspond here to the the quantum
critical point, where the disordered phase (orbital liquid)
emerges from two different types of hidden order, and the first
order transition takes place. A similar transition was shown as 
well by an exact solution of the compass ladder \cite{Brz09}.

The compass model is currently under discussion also due to its
interdisciplinary character. It can be derived using the symmetry
arguments which are necessary for the realization of doubly
degenerate states which are protected from external perturbations
in a wide class of Hamiltonians \cite{Dou05}. The Hamiltonian with
this symmetry can be physically implemented in Josephson
junctions, and it was argued that these junctions provide fault
tolerant quantum bits. Recently magnetic interactions in Mott
insulators with strong spin-orbit coupling were also discussed and
it was pointed out \cite{Jac09} that they may provide a
realization of the exactly solvable Kitaev model on the honeycomb
lattice, which is relevant for quantum computation \cite{Kit06}.

\section{Spin-orbital superexchange}
\label{sec:som}

In several transition metal oxides with active degrees of freedom
one finds coexisting magnetic and orbital order, both in Mott and
in charge transfer insulators. Experimental observations give
frequently some unexpected properties and the question arises how
to explain particular types of observed coexisting spin-orbital
order. First of all, magnetic interactions in these systems
frequently break the cubic symmetry of the perovskite lattice and
AF phases arise with ferromagnetic (FM) interactions along certain
directions (in contrast to the $G$-AF phase with isotropic AF
interactions, the same along each cubic direction). Two of them are 
quite common: ($i$) the $C$-AF phase with FM interactions along the 
$c$ axis as in LaVO$_3$, and ($ii$) the $A$-AF phase with FM
interactions within the $ab$ planes as in LaMnO$_3$ \cite{Tok00}.
These phases follow from the microscopic models (see below) which
justify the complementary behavior of the observed (orbital and 
magnetic) order postulated by Goodenough in the manganites 
\cite{Goo55}:
alternating orbital (AO) order supports FM spin order, while
ferro-orbital (FO) order supports AF spin order. The structure of
spin-orbital superexchange described below allows to understand
better the physical mechanism beyond this complementarity, known
since long as the Goodenough-Kanamori rules \cite{Kan59}.

Realistic superexchange models for transition metal oxides with
orbital degrees of freedom contain both spin and orbital operators.
They may be derived by considering intersite charge excitations in a
Mott insulator in a way similar to the derivation of the $t$--$J$ model
from the Hubbard model (and applying to the cuprates), as described in
Sec. \ref{sec:cup}. This task is somewhat involved, so usually one
considers an effective model with hopping elements between $d$ orbitals
of transition metal ions derived from electron transitions over the
intermediate oxygen orbitals \cite{Zaa93}, as in Eq. (\ref{H0}). The
energy scale for the hopping is set by the largest hopping element $t$:
the $(dd\sigma)$ element in case of $e_g$ systems, and
the $(dd\pi)$ element when only $\pi$ electrons contribute in systems
with degenerate and partly filled $t_{2g}$ orbitals. For noninteracting
electrons the Hamiltonian $H_{0}$ (with crystal-field terms) would lead
to tight-binding bands, but in a Mott insulator one is in
the regime of large Coulomb interaction $U\gg t$, so charge
fluctuations are suppressed and the hopping elements can only
contribute via virtual excitations, leading to the superexchange
as described below.

The derivation of the superexchange involves virtual charge excitations
$d_i^md_j^m\rightleftharpoons d_i^{m+1}d_j^{m-1}$ between two 
neighboring transition metal ions with $m$ $3d$ electrons each, which 
have to be included with the correct excitation energies following from
the structure of local Coulomb interactions in degenerate $d$ states
(a similar expression can also be written for $f$ electrons). These
interactions are well known since long \cite{Ole83}, but nevertheless
simplified expressions can be still found in the literature which in
some cases lead to misleading or even qualitatively incorrect 
conclusions. When only one symmetry class of electrons is involved in 
charge excitations, either $e_g$ or $t_{2g}$ electrons in a perovskite 
system, the interactions read:
\begin{eqnarray}
\label{Heegen}
H_{int}&=&
   U\sum_{i\alpha}n_{i\alpha  \uparrow}n_{i\alpha\downarrow}
+\Big(U'-\frac{1}{2}J_H\Big)\sum_{i,\alpha<\beta}
                    n_{i\alpha}n_{i\beta}                 \nonumber \\
&+& J_H\sum_{i,\alpha<\beta}
\Big( d^{\dagger}_{i\alpha\uparrow}d^{\dagger}_{i\alpha\downarrow}
      d^{       }_{i\beta\downarrow}d^{       }_{i\beta\uparrow}
     +d^{\dagger}_{i\beta\uparrow}d^{\dagger}_{i\beta\downarrow}
      d^{       }_{i\alpha\downarrow}d^{       }_{i\alpha\uparrow}\Big)
                                                          \nonumber \\
&-&2J_H\sum_{i,\alpha<\beta}
    \textbf{S}_{i\alpha}\cdot\textbf{S}_{i\beta}\,,
\end{eqnarray}
with $\bar{\sigma}=-\sigma$. The parameters $\{U,J_H\}$ (with 
$U'=U-2J_H$) determine the excitation energies by the corresponding 
eigenenergies of the local electron-electron Coulomb interaction 
Hamiltonian (\ref{Heegen}). The eigenstates have to be derived for each 
particular situation separately. For instance, if two-hole $d^8$ 
excited states are considered in the 
$d_i^9d_j^9\rightleftharpoons d_i^8d_j^{10}$ transitions for the 
KCuF$_3$ cuprate, there are triplet ($S=1$) and singlet ($S=0$)
states when two holes in the $d^8$ configuration occupy different
orbitals, with the energies $U-3J_H$ and $U-J_H$, as well as two other
(intraorbital) singlet states with the energies $U-J_H$ and $U+J_H$ 
\cite{Ole00}. Note that a double occupancy of either $e_g$ orbital 
(with energy $U$), considered in the context of the intraorbital 
Coulomb interaction (\ref{U}) is {\it not\/} an eigenstate of the local 
Hamiltonian (\ref{Heegen}).

In a general case the interorbital interactions, Coulomb 
$U_{\alpha\beta}$ and exchange $J_{\alpha\beta}$ elements replace $U'$
and $J_H$ --- in contrast to the intraorbital ones (\ref{U}) they 
are anisotropic, but satisfy a constraint which guarantees
the invariance of interactions in the orbital space \cite{Ole84},
\begin{equation}
\label{Uab}
  U=U_{\alpha\beta}+2J_{\alpha\beta}\,,
\end{equation}
for each pair of interacting orbitals $\{\alpha\beta\}$. These
interactions are frequently parametrized by the Racah parameters
$\{A,B,C\}$ \cite{Gri71}, and one finds that
\begin{equation}
\label{U}
  U=A+4B+3C\,,                                                   \\
\end{equation}
and is identical for all $3d$ orbitals, while Hund's exchange depends on
the orbital states of the pair of interacting electrons --- for $e_g$
electrons,
\begin{equation}
\label{JHe}
  J_H^e=4B+C\,,                                                   \\
\end{equation}
and it is somewhat stronger than for $t_{2g}$ electrons,
\begin{equation}
\label{JHt}
  J_H^t=3B+C\,.                                                   \\
\end{equation}
Thus the parameter $J_H$ used in Eq. (\ref{Heegen}) refers to the
above values depending on whether a system with $e_g$ or $t_{2g}$
orbital degrees of freedom is
considered. More details about the structure of local Coulomb
interactions which depends on a single parameter,
\begin{equation}
\label{eta}
  \eta=\frac{J_H}{U}\,,                                                   \\
\end{equation}
may be found in Ref. \cite{Ole84}, while the experimental
values of the Racah parameters $\{B,C\}$  which are known with high
accuracy from the atomic spectra are given in Ref. \cite{Zaa90}
for several transition metal ions. Unfortunately, the value of $U$
(or $A$) is known only with much lower accuracy and the mechanism of
screening which leads to the values $U\sim 5-10$ eV is difficult to
implement in the theory. Hence, the value of $U$ is frequently used as
a parameter, unless it can be derived from the experimental data, as
for instance from the optical excitations, see below.

The above structure of the electron-electron interactions (\ref{Heegen})
determines the excitation energies $\varepsilon_n$ due to the multiplet 
states in charge transitions 
$d_i^md_j^m\rightleftharpoons d_i^{m+1}d_j^{m-1}$
which enter the superexchange in the respective denominators of
$4t^2/\varepsilon_n$. Examples of such spectra are presented in Ref.
\cite{Ole05}. As a rule, the high-spin states have the lowest energy
$U-3J_H$ independently of the electron number $m$ in the electronic
configuration $d^m$ under consideration, while the energies of
low-spin states depend on $m$, and may even contain fractions of
$J_H$ due to the anisotropy of $J_{\alpha\beta}$ Hund's elements,
as found for instance in the case of LaMnO$_3$ \cite{Fei99}.

The spin-orbital superexchange is the effective low-energy Hamiltonian
which involves products of spin and orbital operators. The spin 
interactions are described by spin scalar products 
${\vec S}_i\cdot{\vec S}_j$ on each bond $\langle ij\rangle$ connecting 
two nearest neighbor transition metal ions and obey the SU(2) symmetry,
while the orbital operators $\{{\vec\tau}_i,{\vec\tau}_j\}$,
with ${\vec\tau}_i=\{\tau_i^+,\tau_i^-,\tau_i^z\}$,
obey only much lower symmetry (at most cubic for a cubic lattice) and
appear either as a scalar product ${\vec\tau}_i\cdot{\vec\tau}_j$, or
only as certain components, for instance an Ising
term $\tau_i^z\tau_j^z$, resembling the interactions in the 2D compass
model (\ref{com2D}). In general the spin-orbital superexchange is of the
form \cite{Ole05},
\begin{equation}
{\cal H}_J = J \sum_{\langle ij \rangle \parallel\gamma} \left\{
{\hat {\cal J}}_{ij}^{(\gamma)} \left( {\vec S}_i \cdot {\vec S}_j
+S^2\right) + {\hat {\cal K}}_{ij}^{(\gamma)} \right\}\,,
\label{som}
\end{equation}
with the constant $J$ defined in Eq.  (\ref{J}).
The orbital operators ${\hat{\cal J}}_{ij}^{(\gamma)}$ and
${\hat{\cal K}}_{ij}^{(\gamma)}$ depend on the direction $\gamma=a,b,c$
in the cubic lattice and involve the active orbitals on each
bond $\langle ij \rangle$ (either $e_{g}$ or $t_{2g}$) along  direction
$\gamma$ --- they participate in $d^m_i d^m_j \rightleftharpoons
d^{m+1}_i d^{m-1}_j$ virtual excitations, and thus these interactions 
have the symmetry of the lattice ({\it i.e.}~cubic symmetry in the 
perovskites). As an example we introduce here
the superexchange between $V^{3+}$ ions in the $d^2$
configuration with $S=1$ spins \cite{Kha01}, as realized in $R$VO$_3$
perovskites considered below in Sec. \ref{sec:van} and controlled by
the orbital operators:
\begin{eqnarray}
\label{orbj}
{\hat J}_{ij}^{(\gamma)}\!\!&=&\!\!
\frac{1}{2}\left\{(1+2\eta r_1)
\left({\vec\tau}_i\cdot {\vec\tau}_j
     +\frac{1}{4}n_i^{}n_j^{}\right)\right.          \nonumber \\
&-&\! \left.\eta r_3
    \left({\vec\tau}_i\times{\vec\tau}_j+\frac{1}{4}n_i^{}n_j^{}\right)
-\frac{1}{2}\eta r_1(n_i+n_j)\right\}^{(\gamma)}\,,             \\
\label{orbk}
{\hat K}_{ij}^{(\gamma)}\!\!&=&\!\!
\left\{\eta r_1
\left({\vec\tau}_i\cdot {\vec\tau}_j+\frac{1}{4}n_i^{}n_j^{}\right)
 +\eta r_3\left({\vec\tau}_i\times {\vec\tau}_j
             +\frac{1}{4}n_i^{}n_j^{}\right)\right.  \nonumber \\
&-&\!\left.
   \frac{1}{4}(1+\eta r_1)(n_i+n_j)\right\}^{(\gamma)}\,.
\end{eqnarray}
They arise from the
$d_i^2d_j^2\rightleftharpoons d_i^3d_j^1$ charge excitations,
leading either to high-spin or to low-spin $d^3_i$ configurations,
so Hund's exchange in the
multiplet structure of a V$^{2+}$ ions enters via the coefficients
$r_1=1/(1-3\eta)$ and $r_3=1/(1+2\eta)$ (the low-spin excitations 
occur also at energy $U$, so the corresponding coefficient is
$r_2\equiv 1$). Eqs. (\ref{orbj}) and (\ref{orbk}) are general and 
refer to two active orbital flavors along the cubic axis $\gamma$.
The leading orbital interactions are proportional to the scalar
products $({\vec\tau}_i\cdot {\vec\tau}_j)^{(\gamma)}$ of orbital
operators on the bonds as both orbitals are active and may generate
charge excitations, but the structure of local Coulomb interactions 
(\ref{Heegen}) is responsible for additional terms,
\begin{equation}
\left({\vec\tau}_i\times {\vec\tau}_j\right)^{(c)}=
\frac{1}{2}\left(\tau_i^+\tau_j^++\tau_i^-\tau_j^-\right)
+\tau_i^z\tau_j^z\,,
\end{equation}
which violate the conservation of the orbital quantum numbers.
The operator $n_i^{(\gamma)}$ stands for the number of active
electrons at site $i$ along the bond $\langle ij\rangle$, for
instance for a bond along the $c$ axis this number is
$n_i^{(c)}=n_{ia}^{}+n_{ib}^{}$ [this notation for the $t_{2g}$
orbitals is defined in Eq. (\ref{t2g})].

The superexchange model (\ref{som}) consists typically of several
terms which originate from different charge excitations. This
feature made it possible to relate the averages of these different
excitations to the spectral weights in the optical spectroscopy
\cite{Kha04}, and serves now as a standard theoretical tool to
explain the observed anisotropy and temperature dependence of the
spectral weights in the optical spectra \cite{Ole05}. In a
correlated insulator the electrons are almost localized and the only 
kinetic energy which is left \cite{Aic02} is associated with the 
same virtual charge excitations that contribute also to the
superexchange (\ref{som}). Therefore, we one may
define the individual kinetic energy contributions
$K_n^{(\gamma)}$, which refer to different energy regimes in the
optical transitions and can be determined from the superexchange
using the Hellman-Feynman theorem \cite{Bae86},
\begin{equation}
\label{hefa}
K_n^{(\gamma)}=-2\big\langle H_n^{(\gamma)}(ij)\big\rangle.
\end{equation}
For convenience, we define the energy contribution $K_n^{(\gamma)}$
for the Hubbard subband $n$ as a positive quantity.

The magnetic properties of the transition metal oxides with active
orbital degrees of freedom are usually discussed in terms of magnetic
exchange constants which determine both the type of the magnetic
order in the ground state (at $T=0$) and the magnetic excitations
(magnons observed in the neutron scattering experiments). The exchange
constants are usually found for a bond $\langle ij\rangle$ along each
nonequivalent axis $\gamma$ by
averaging over the orbital operators in Eq. (\ref{som}),
\begin{equation}
J_{ij}=\langle {\hat{\cal J}}_{ij}^{(\gamma)}\rangle\,,
\label{Jij}
\end{equation}
which leads to an anisotropic spin exchange model Hamiltonian adequate 
for anisotropic magnetic phases, such as for instance  $A$-AF or $C$-AF 
phase realized for instance in LaMnO$_3$ and LaVO$_3$,
\begin{equation}
\label{Hs}
H_s =J_{ab}\sum_{\langle ij\rangle_{ab}}{\vec S}_i\cdot{\vec S}_j
+J_c\sum_{\langle ij\rangle_c}{\vec S}_i\cdot{\vec S}_j\,.
\end{equation}
This procedure assumes implicitly that spin and orbital operators can
be separated from each other and ignores the possibility of quantum
entanglement \cite{Ole06} and of composite spin-orbital excitations
introduced in Refs. \cite{Ole00,Fei98}. It turns out that such
excitations play a prominent role in destabilizing the classical AF
long-range order in the $d^9$ spin-orbital model \cite{Fei97},
and have observable consequences in the perovskite vanadates,
see Sec. \ref{sec:van}.

In some cases, however, the spin and orbital degrees of freedom may be
disentangled and the obtained theoretical results explain well the
experimental findings. One of the best examples is LaMnO$_3$, where
the exchange constants $\{J_{ab},J_c\}$ deduced from the neutron
scattering \cite{Mou96} can be explained by the superexchange model
assuming a classical ansatz for the ground state with AO order,
\begin{equation}
\label{oo}
|\Phi_0\rangle=\prod_{i\in A}|\theta_A\rangle_i
               \prod_{j\in B}|\theta_B\rangle_j,
\end{equation}
with the orbital states, $|\theta_A\rangle_i$ and $|\theta_B\rangle_j$,
characterized by opposite angles ($\theta_A=-\theta_B$) on two
sublattices $A$ and $B$ in the $ab$ planes, and repeated in the
subsequent planes along the $c$ axis. The AO order is stable below the 
orbital transition temperature $T_{\rm OO}\simeq 760$ K \cite{Goo06}, 
which is rather high compared with the N\'eel temperature
$T_N\simeq 140$ K --- therefore one may consider the AO order
between the sublattices ($i\in A$, $j\in B$),
\begin{eqnarray}
\label{ood9}
|\theta_A\rangle_i&=&\cos\left(\frac{\theta}{2}\right)|z\rangle_i
                    +\sin\left(\frac{\theta}{2}\right)|x\rangle_i\,,
\nonumber\\
|\theta_B\rangle_j&=&\cos\left(\frac{\theta}{2}\right)|z\rangle_j
                    -\sin\left(\frac{\theta}{2}\right)|x\rangle_j\,.
\end{eqnarray}
as frozen in the temperature range $T<300$ K relevant for the magnetic
excitations. Here we used the simplified notation for the $e_g$
orbital basis introduced in Eq. (\ref{eg}).
Using the well motivated parameter set, the experimental values of
the exchange constants in LaMnO$_3$ are reproduced by the angle
$\theta\simeq 94^{\circ}$, but it has been shown that somewhat
higher values of $\theta$ would also be consistent with a model
including explicitly charge transfer superexchange terms \cite{Ole05}.
In any case, the occupied orbitals in LaMnO$_3$ are closer to
symmetric/antisymmetric combinations of $\{|z\rangle,|x\rangle\}$ than
to the directional $3x^2-r^2/3y^2-r^2$ orbitals, as illustrated in the
early literature on the subject.

The experimental proof that the spin and orbital operators may be
{\it disentangled\/} in LaMnO$_3$ is provided by the optical
spectroscopy, which shows rather distinct anisotropy of the low-energy
spectral weights between the polarization in $ab$ planes on the one 
hand and along the $c$ axis on the other \cite{Kov04}. 
It is quite remarkable that the
temperature dependence of these spectral weights may be very well
explained by the spin-orbital superexchange model of Ref. \cite{Fei99},
using the same parameters as those used to calculate the exchange
constants, in the broad temperature range $0<T<300$ K \cite{Kov04}.
It follows alone from the temperature variation of spin correlation
functions for a bond $\langle ij\rangle$ within the $ab$ planes and 
along the $c$ axis,
\begin{equation}
\label{spins}
s_{ab}=\langle\vec{S}_i\cdot\vec{S}_j\rangle^{(ab)}\,, \hskip 1cm
   s_c=\langle\vec{S}_i\cdot\vec{S}_j\rangle^{(c)}\,,
\end{equation}
while
the AO order remains unchanged, as explained in Ref. \cite{Ole05}.

In spite of the disentangled spin and orbital dynamics in the $R$MnO$_3$
perovskites, where $R$=Lu,$\cdots$,La stands for a rare earth atom,
their several properties are not fully understood. One of
them is the nature of the insulating state which comes partly due to JT
interactions \cite{Ben99,Hel06,Mil07} and partly due to the orbital
superexchange interactions \cite{Fei99}. Another puzzling feature
is the phase diagram of the $R$MnO$_3$ family of compounds, where
$R$=Lu,$\cdots$,La stands for a rare earth atom, which exhibits a
phase transition from the $A$-AF to a rather peculiar $E$-AF phase
\cite{Goo06}.

\section{Fingerprints of spin-orbital entanglement in the 
$R$VO$_3$ perovskites}
\label{sec:van}

\subsection{Spin-orbital entanglement}
\label{sec:enta}

The coupling between spin and orbital operators in the spin-orbital
superexchange may be quite strong in some cases --- the excellent
example of this coupling are the vanadium perovskites, see below.
Although the $C$-AF phase observed in the
entire family of $R$VO$_3$ compounds \cite{Miy03,Miy06}, where
$R$=Lu,$\cdots$,La stands for a rare earth atom, satisfies to some
extent the Goodenough--Kanamori rules \cite{Kan59}, with FM order
along the $c$ axis where the active $a$ and $b$ orbitals (\ref{t2g})
alternate --- the AO order is very weak here and the orbital
fluctuations play a very important role \cite{Kha01}. This situation is
opposite to the frozen AO order in LaMnO$_3$, which can explain both the
observed magnetic exchange constants and the distribution of the optical
spectral weights. In LaVO$_3$ the FM exchange interaction is enhanced
far beyond the usual mechanism following from the splitting between the
high-spin and low-spin states due to finite Hund's exchange $J_H$.
Evidence
of orbital fluctuations in the $R$VO$_3$ perovskites was also found in
pressure experiments, which show a distinct competition between the
$C$-AF and $G$-AF spin order, accompanied by the complementary $G$-AO
and $C$-AO order of $\{a,b\}$ orbitals \cite{Miy06}.

To understand better the essence of entangled spin-orbital states, we
present first the results of the model calculation with four-site
chains along the $c$ axis, described by the spin-orbital superexchange
models relevant for titanates and vanadates. These calculations served
to identify spin-orbital entangled states for increasing multiplet
splitting $\propto\eta$ \cite{Ole06}. A prototype model to study
frustration and entanglement in coupled spin and pseudospin
(orbital) systems is the one--dimensional (1D) SU(4) model \cite{Li98}.
This example is remarkable, as in a purely spin 1D model one expects no
frustration when only nearest-neighbor interactions are present.
However, both spins and pseudospins appear here on a completely
symmetrical and equal footing with joint spin-pseudospin operators and
compete with each other, forming a group of elementary generators in
the SU(4) symmetry. Three types of elementary excitations contribute to
the thermodynamic properties: spin, orbital, and joint spin-orbital ones.
This is indeed confirmed by the entropy data of this model obtained from
a numerical analysis \cite{Fri99}, which increases three times faster
than that of the 1D AF Heisenberg model. This also implies that the
intersite correlation functions are intimately interrelated and it is
impossible to separate the two subsystems. Hence, one has to
treat explicitly entangled spin-pseudospin states.

As a useful tool to verify the Goodenough-Kanamori rules \cite{Kan59}
in spin-orbital models for $t_{2g}$ electrons with spins either $S=1$
or $S=1/2$ and pseudospins $\tau=1/2$,
we introduce spin and orbital correlations defined for a bond
$\langle ij\rangle$,
\begin{equation}
\label{sij}
S_{ij}=\langle{\vec S}_i\cdot{\vec S}_j\rangle/(2S)^2\,, \hskip .7cm
T_{ij}=\langle{\vec T}_i\cdot{\vec T}_j\rangle\,.
\end{equation}
When they are compared with each other, and with the composite
spin-orbital correlation function defined as a difference between
the exact value and the MF factorized correlations on a bond
$\langle ij\rangle$ \cite{Ole06},
\begin{equation}
\label{cij}
C_{ij}=
\left\{\big\langle({\vec S}_i\cdot{\vec S}_j)
                ({\vec T}_i\cdot{\vec T}_j)\big\rangle
    -\big\langle {\vec S}_i\cdot{\vec S}_j \big\rangle
     \big\langle {\vec T}_i\cdot{\vec T}_j \big\rangle\right\}/(2S)^2\,,
\end{equation}
one may conclude whether the spin and orbital operators are
disentangled. If $C_{ij}=0$, the spin and orbital operators are
disentangled and their MF decoupling is exact, while if $C_{ij}<0$
--- spin and orbital operators are entangled, and joint
spin-orbital fluctuations contribute even at $T=0$. Two
spin-orbital models were investigated in Ref. \cite{Ole06}: ($i$)
the titanate model for $d^1$ ionic configurations of Ti$^{3+}$
ions in the $R$TiO$_3$ perovskites with $S=1/2$ \cite{Kha00}, and
($ii$) the vanadate model for $d^2$ configurations of V$^{3+}$
ions in the $R$VO$_3$ perovskites with $S=1$ \cite{Kha01}. For
more details about the structure of the superexchange ${\cal H}_J$
(\ref{som}) in both models see for instance Ref. \cite{Ole05}.

As the chain-like cluster is 1D and only two orbital $\{a,b\}$ flavors
contribute in each case, one recovers the SU(4) model in the $d^1$ 
(titanate) case  at $\eta=0$, and $S_{ij}=T_{ij}=C_{ij}=-0.25$ for 
$N=4$ sites \cite{Ole05}. By a closer
inspection one finds that the ground state wave function for the
four-site cluster is close to a total spin-orbital singlet,
involving a linear combination of
(spin singlet/orbital triplet) and
(spin triplet/orbital singlet) states for each bond $\langle ij\rangle$.
This result manifestly contradicts the celebrated Goodenough-Kanamori 
rules \cite{Kan59}, as both spin and orbital correlations have the same 
sign. At finite $\eta$ the SU(4) degeneracy of all intersite 
correlations is removed --- one finds $T_{ij}<C_{ij}<S_{ij}<0$ in the 
regime of spin singlet ($S=0$) ground state, and the Goodenough-Kanamori 
rule with complementary spin/orbital correlations is still violated. 
A qualitatively similar case is found in a mathematical 
SU(2)$\otimes$SU(2) model (not realized in transition metal oxides) 
\cite{Aff00}, where the ground state is entangled in a
broad range of parameters, including the exactly solvable case with
alternating spin and orbital singlets on the bonds \cite{Pss07}.

The vanadate $d^2$ model behaves also in a similar way in a range of
small values of $\eta$, with all three $S_{ij}$, $T_{ij}$ and $C_{ij}$
correlations being negative. Again, the composite spin-orbital 
correlations are here finite ($C_{ij}<0$), spin and orbital variables 
are {\it entangled\/}, and the MF factorization of the ground state 
into spin and orbital part fails. Only for sufficiently large $\eta$ do 
the spins reorient in the FM ground state, and decouple from the 
orbitals. In this regime, corresponding to the experimentally observed 
$C$-AF phase of LaVO$_3$ (and other cubic vanadates), spin-orbital
entanglement ceases to exist in the ground state. However, as we will
see below, it has still remarkable consequences at finite temperature, 
where entangled spin-orbital states again play a role.

A crucial observation concerning the applicability of the
Goodenough-Kanamori rules to the quantum models of $t_{2g}$ electrons
in one dimension can be made by comparing spin exchange constants
$J_{ij}$ calculated from Eq. (\ref{Jij}) with the actual values of
intersite spin correlations $S_{ij}$ (\ref{sij}). One finds that
exchange interaction is negative ($J_{ij}<0$), so formally favors FM
spin orientation, in the singlet phase at low values of $\eta$, but it
is accompanied by AF spin correlations ($S_{ij}<0$). This demonstrates
that the ground state energy calculated in the MF theory would be
{\it enhanced\/}, so the MF approach cannot be used \cite{Ole06}. This 
result follows from large spin-orbital fluctuations which cause also 
large fluctuations of the exchange constants around the average value,
measured by $\delta J=
\{\langle ({\hat{\cal J}}_{ij}^{(\gamma)})^2\rangle-J_{ij}^2\}^{1/2}$.
Altogether, this result challenges the usual interpretation of the
magnetic data in the spin-orbital systems with the exchange constants
determined by averaging over the orbital operators, see Eq. (\ref{Jij}).
Fortunately, Hund's exchange is large enough in real materials and this
conceptual difficulty is removed in transition metal oxides, but one 
may expect that experimental results in the range of finite temperature 
will depend on the entangled states discussed above.

\subsection{Phase diagram of the $R$VO$_3$ perovskites}

The phase diagram of the $R$VO$_3$ perovskites \cite{Miy03,Miy06} is
qualitatively different from the one for the $R$VO$_3$ perovskites
\cite{Goo06} and indicates the proximity of spin and orbital energy
scales. Experimental studies have shown that the $C$-AF order is
common to the entire family of the $R$VO$_3$ vanadates, and in general
the magnetic transition occurs below the orbital transition,
$T_{N1}<T_{\rm OO}$, except for LaVO$_3$ with $T_{N1}\simeq T_{\rm OO}$ 
\cite{Miy03,Miy06}. When the ionic radius $r_R$ decreases, the N\'eel 
temperature $T_{N1}$ also decreases, while the orbital transition 
temperature $T_{\rm OO}$ increases, passes through a
maximum close to YVO$_3$, and next decreases towards LuVO$_3$. This
provided an experimental challenge to the theory which was addressed
only recently using the spin-orbital superexchange model \cite{Hor08}.
One finds that the $C$-AF order develops in LaVO$_3$ below
$T_{N1}\simeq 143$ K, and is almost immediately followed by a weak
structural transition stabilizing the weak $G$-AO order at
$T_{\rm OO}\simeq 141$ K \cite{Miy03,Miy06}. This provides a constraint
on the theoretical model. Remarkably, the magnetic
order parameter in the $C$-AF phase of LaVO$_3$ is strongly reduced to
$\simeq 1.3\mu_B$, much below the reduction expected from quantum
fluctuations in the $C$-AF phase (being only 6\% for $S=1$ spins
\cite{Rac02}) --- also this reduction of the
measured magnetization could not be explained so far.

In order to unravel the physical mechanism responsible for the decrease
of $T_{\rm OO}$ from YVO$_3$ to LuVO$_3$ one has to analyze in more
detail the evolution of GdFeO$_3$ distortions with for decreasing ionic
radius $r_R$ \cite{Hor08}. Such distortions are common for the
perovskites \cite{Pav05}, and one expects that they should increase
when the ionic radius $r_R$ decreases, as observed in the $R$MnO$_3$
perovskites \cite{Goo06}. In the $R$VO$_3$ family the distortions
are described by two subsequent rotations of VO$_6$ octahedra:
($i$) by an angle $\vartheta$ around the $b$ axis, and
($ii$) by an angle $\varphi$ around the $c$ axis.
Increasing angle $\vartheta$ causes a decrease of
V--O--V bond angle along the $c$ direction, being $\pi-2\vartheta$,
and leads to an orthorhombic lattice distortion $u=(b-a)/a$, where
$a$ and $b$ are the lattice parameters of the $Pbnm$ structure of
$R$VO$_3$. By the analysis of the structural data for the $R$VO$_3$
perovskites \cite{Ree06,Sag06} one finds the following empirical
relation between the ionic radius $r_R$ and the angle $\vartheta$:
\begin{equation}
\label{r}
r_R=r_0-\alpha\sin^2\vartheta\,,
\end{equation}
where $r_0=1.5$ \AA{} and $\alpha=0.95$ \AA{} are the empirical
parameters. This allows one to use the angle $\vartheta$ to
parametrize the dependence of the microscopic parameters of the
Hamiltonian and to investigate the transition temperatures
$T_{\rm OO}$ and $T_{N1}$ as functions of $r_R$.

The spin-orbital model introduced in Ref. \cite{Hor08} to describe
the phase diagram of $R$VO$_3$ reads:
\begin{eqnarray}
\label{som2}
{\cal H}\!\!\!&=&\!\!\!J\!\!\sum_{\langle ij\rangle\parallel\gamma}\!
\left\{\!\Big({\vec S}_i\!\cdot\!{\vec S}_j\!+\!S^2\Big){{\cal
J}}_{ij}^{(\gamma)}
+ {{\cal K}}_{ij}^{(\gamma)}\!\right\}
+E_z(\vartheta)\!\sum_i\!e^{i{\vec R}_i{\vec Q}}\tau_i^z
\nonumber \\
&-& V_{c}(\vartheta)\!\!\sum_{\langle ij\rangle\parallel c}\!\!
\tau_i^z\tau_j^z
+V_{ab}(\vartheta)\!\sum_{\langle ij\rangle\parallel ab}\tau_i^z\tau_j^z
\nonumber \\
&-&gu\sum_i\tau_i^x+\frac12 N K(u-u_0(\vartheta))^2\,,
\end{eqnarray}
where $\gamma=a,b,c$ labels the cubic axes, and the operators are
given by Eqs. (\ref{orbj}) and (\ref{orbk}). The superexchange is
supplemented by the crystal field term $\propto E_z$, the orbital
interactions terms $\propto V_c$ and $\propto V_{ab}$ induced by lattice
distortions, and the orbital-lattice term $\propto g$ which is
counteracted by the lattice elastic energy $\propto K$. All these terms
are necessary in a realistic model which reproduces the behavior of the
$R$VO$_3$ perovskites at finite temperature.

The crystal
field splitting breaks the cubic symmetry in distorted VO$_6$ octahedra,
as obtained in the electronic structure calculations \cite{And07} and
from the point charge model \cite{Hor08}, and the actual filling of
$t_{2g}$ orbitals is:
\begin{equation}
n_{ic}=1,\hskip .7cm n_{ia}+n_{ib}=1\,,
\end{equation}
so the superexchange (\ref{orbj}) and (\ref{orbk}) in Eq. (\ref{som2})
is expressed by the orbital operators 
${\vec\tau}_i=\{\tau_i^+,\tau_i^-,\tau_i^z\}$ (and their components) as 
explained in Sec. IV.
The splitting $\propto E_z$ between $a$ ($yz$) and $b$ ($zx$) orbitals 
is given by the pseudospin $\tau_i^z$ operators,
\begin{equation}
\tau_i^z=\frac{1}{2}(n_{ia}-n_{ib})\,,
\end{equation}
which refer to two active orbital flavors $\{a,b\}$ in $R$VO$_3$.
It is characterized by the vector ${\vec Q}=(\pi,\pi,0)$ in reciprocal 
space --- it alternates in the $ab$
planes, but is uniform along the $c$ axis. Thus, this splitting
competes with the (weak) $G$-AO order
supporting the observed $C$-AF phase at temperature $T<T_{N1}$.

In addition, the model (\ref{som2}) includes: ($i$) intersite
orbital interactions $\propto V_{ab},V_c$ (which originate from
the coupling to the lattice), and ($ii$) orbital-lattice term
$\propto g$ which induces orbital polarization
$\langle\tau_i^x\rangle\neq 0$ when the lattice distortion $u$
increases. The orbital interactions induced by the distortions of
the VO$_6$ octahedra and by GdFeO$_3$ distortions of the lattice,
$V_{ab}>0$ and $V_c>0$, also favor the $C$-AO order (like the
crystal field $E_z>0$). Note that $V_c>0$ counteracts the orbital
interactions included in the superexchange via ${\hat K}_{ij}^{(c)}$ 
operators (\ref{orbk}). The last two terms in Eq. (\ref{som2}) describe 
the  linear coupling $\propto g>0$ between active $\{yz,zx\}$ orbitals
and the orthorhombic lattice distortion $u$. The elastic energy
which counteracts lattice distortion $u$ is given the force
constant $K$, and $N$ is the number of $V^{3+}$ ions. The coupling
$\propto gu$ acts as a transverse field in the pseudospin space.
While the eigenstates $\frac{1}{\sqrt{2}}(|a\rangle\pm|b\rangle)$
favored by $\tau_i^x$ cannot be realized due to the competition
with all the other terms, increasing lattice distortion $u$
(increasing angle $\vartheta$) modifies the orbital order and
intersite orbital correlations.

The crystal field splitting $E_z(\vartheta)$, orbital interactions
$\{V_{ab}(\vartheta),V_c(\vartheta)\}$, and the orbital-lattice
coupling $g_{\rm eff}(\vartheta)\equiv gu$ depend on the tilting angle
$\vartheta$. In case of $V_{c}$ one may argue that its dependence on
the angle $\vartheta$ is weak, and a constant
$V_{c}(\vartheta)\equiv 0.26J$ was chosen in Ref. \cite{Hor08} in order
to satisfy the experimental constraint that the $C$-AF and $G$-AO order
appears almost simultaneously in LaVO$_3$ \cite{Miy03}. The
experimental value $T_{N1}^{\rm exp}=143$ K for LaVO$_3$ \cite{Miy03}
was fairly well reproduced in the present model taking $J=200$ K.
The functional dependence of the remaining two parameters
$\{E_z(\vartheta),V_{ab}(\vartheta)\}$ on the tilting angle $\vartheta$
was derived from the point charge model \cite{Hor08} using the
structural data for the $R$VO$_3$ series \cite{Ree06,Sag06} ---
one finds:
\begin{eqnarray}
\label{Ez}
E_z(\vartheta)&=&J\,v_z\,\sin^3\vartheta\cos\vartheta\,, \\
\label{vab}
V_{ab}(\vartheta)&=&J\,v_{ab}\,\sin^3\vartheta\cos\vartheta\,.
\end{eqnarray}
Finally, the effective coupling to the lattice distortion has to
increase faster with the increasing angle  $\vartheta$, and the 
following dependence was shown \cite{Hor08} to give a satisfactory 
description of the phase diagram of the $R$VO$_3$ perovskites:
\begin{equation}
\label{geff}
g_{\rm eff}(\vartheta)=J\,v_{g}\,\sin^5\vartheta\cos\vartheta\,.
\end{equation}
Altogether, magnetic and orbital correlations described by the
spin-orbital model (\ref{som}), and the
magnetic $T_{N1}$ and orbital $T_{\rm OO}$ transition temperatures
depend on three parameters: $\{v_z,v_{ab},v_g\}$.

Due to the proximity of both orbital and magnetic phase
transitions in the $R$VO$_3$ perovskites, it is crucial to design
the MF approach in such a way that the spin-orbital coupling is
described {\it beyond\/} the factorization of spin and orbital
operators. On the one hand, the correct MF treatment of the
orbital and magnetic phase transitions in the $R$VO$_3$ vanadates
requires the coupling between the on-site orbital,
$\langle\tau^z\rangle_G\equiv\frac12|\langle\tau^z_i-\tau^z_j\rangle|$,
and spin order parameters in the $C$-AF phase, $\langle
S_i^z\rangle_C$, as well as a composite $\langle
S_i^z\tau_i^z\rangle$ order parameter, similar to that used for the 
$R$MnO$_3$ perovskites \cite{Fei99}. On the other hand, the on-site MF 
theory including the above coupling \cite{Sil03} does not suffice for 
the $R$VO$_3$ compounds as the orbital singlet correlations
$\langle{\vec\tau}_i\cdot{\vec\tau}_j\rangle$ on the bonds
$\langle ij\rangle$ along the $c$ axis play so crucial role in
stabilizing the $C$-AF phase \cite{Kha01} and the orbital
fluctuations are important \cite{God07}. Therefore, the minimal
physically acceptable approach to the present problem is a
self-consistent calculation for a bond $\langle ij\rangle$ along
the $c$ axis, coupled by the MF terms to its neighbors along all
three cubic axes \cite{Hor08}. This procedure, with properly
selected model parameters, was shown to be successful in
reproducing the experimental phase diagram of Ref. \cite{Miy06}.
One finds that indeed the orbital order occurs below a higher
temperature than the magnetic one in the $R$VO$_3$ perovskites to
the left from LaVO$_3$, i.e. with smaller ionic radius $r_R$.

As presented in Ref. \cite{Hor08}, the remarkable dependence of both
spin $T_{N1}$ and orbital $T_{\rm OO}$ transition temperature in the
$R$VO$_3$ perovskites follows from the respective changes in the
orbital correlations with decreasing $r_R$. First, the singlet
correlations are drastically suppressed from LaVO$_3$ towards LuVO$_3$.
Second, the increase of orbital intersite interactions due to the JT
term (\ref{vab}), induces steady increase of the orbital temperature
$T_{\rm OO}$ with decreasing $r_R$. Finally,
while $\langle\tau^x_i\rangle\simeq 0.03$ is rather weak in
LaVO$_3$, it steadily increases along the $R$VO$_3$ perovskites when
$r_R$ decreases, and finally it becomes as important as the orbital
order parameter itself, i.e.
$\langle\tau^x_i\rangle\simeq\langle\tau^z_i\rangle_G$. Note that in 
the entire parameter range the latter order parameter is substantially
reduced from the classical value $\langle\tau^z\rangle_{G,\rm max}=
\frac12$ by singlet orbital fluctuations in the entire parameter 
regime, being $\langle\tau^z_i\rangle_G\simeq 0.32$
and 0.36 for LaVO$_3$ and LuVO$_3$, respectively.

It is quite remarkable that the above changes in the orbital state
modify the magnetic exchange constants $\{J_{ab},J_c\}$ along both
nonequivalent cubic directions, see Eq.  (\ref{Jij}), and thus the
value of $T_{N1}$ is reduced with decreasing $r_R$. Note that the
superexchange energy $J$ does not change, so the entire effect stems
from the orbital correlations \cite{Hor08}. This also implies that the
width of the magnon band given at $T=0$ by
$W_{C-{\rm AF}}=4(J_{ab}+|J_c|)$ is reduced by a factor
close to 1.8 from LaVO$_3$ to YVO$_3$, in agreement with surprisingly
low magnon energies observed in the $C$-AF phase of YVO$_3$
\cite{Ulr03}.

Summarizing, the microscopic model (\ref{som2}) describes gradual
changes of the orbital and magnetic correlations under the
coupling to the lattice which suppresses orbital fluctuations
generated by virtual charge fluctuations responsible for the
spin-orbital superexchange. It provides an almost quantitative 
understanding of the systematic experimental trends for both orbital 
and magnetic transitions in the $R$VO$_3$ perovskites \cite{Hor08}, 
and is able to reproduce the observed non-monotonic variation of the 
orbital transition temperature $T_{\rm OO}$ for decreasing $r_R$. 
However, the theoretical description of the magnetic transition to 
the $G$-AF phase at $T_{N2}$, which occurs for small $r_R$ 
\cite{Miy03}, remains to be addressed by future theory. More examples 
of spin-orbital entanglement in the field of the perovskite vanadates
are shortly discussed in the next two subsection.

\subsection{Optical spectral weights for LaVO$_3$}

As a second example of spin-orbital entanglement in the cubic vanadates
at finite temperature we discuss briefly the evaluation of the optical 
spectral weights from the spin-orbital superexchange for LaVO$_3$, 
following Eq. (\ref{hefa}). First we rewrite the superexchange operator 
$H^{(\gamma)}(ij)$ for a bond $\langle ij\rangle\parallel\gamma$, 
contributing to operator ${\cal H}_J$ (\ref{som}), as a superposition 
of $d_i^2d_j^2\rightleftharpoons d_i^3d_j^1$ charge excitations to
different spin states in upper Hubbard subbands labelled by $n$
\cite{Kha04},
\begin{equation}
\label{Hn} H^{(\gamma)}(ij)=\sum_n H_{n,ij}^{(\gamma)}\,.
\end{equation}
One finds the superexchange terms $H^{(c)}_{n,ij}$ for a bond
${\langle ij\rangle}$ along the $c$ axis \cite{Kha04},
\begin{eqnarray}
\label{H1c} H_{n,ij}^{(c)}\!\!\!&=&\!\!-\frac{1}{3}Jr_1\! (2\!+\!\vec
S_i\!\cdot\!\vec S_j)\!
\left(\!\frac{1}{4}-\vec \tau_i\cdot\vec \tau_j\right),                  \\
\label{H2c} H_{n,ij}^{(c)}\!\!\!&=&\!\!-\frac{1}{12}J\!(1\!-\!\vec S_i\!\cdot\!\vec
S_j)\! \left(\!\frac{7}{4}-\!\tau_i^z\tau_j^z\!-\!\tau_i^x\tau_j^x\!
+\!5\tau_i^y \tau_j^y\!\right),                                           \\
\label{H3c} H_{n,ij}^{(c)}\!\!\!&=&\!\!-\frac{1}{4}Jr_3\! (1\!-\!\vec
S_i\!\cdot\!\vec S_j)\!
\left(\!\frac{1}{4}+\!\tau_i^z\tau_j^z\!+\!\tau_i^x\tau_j^x\! -\!\tau_i^y
\tau_j^y\!\right),
\end{eqnarray}
and $H^{(ab)}_{n,ij}$ for a bond in the $ab$ plane,
\begin{eqnarray}
\label{H1a} H_{n,ij}^{(ab)}\!\!\!&=&\!\!-\frac{1}{6}Jr_1\!\left(2\!+\!\vec
S_i\!\cdot\!\vec S_j\right)\!
\left(\!\frac{1}{4}-\tau_i^z\tau_j^z\right),                 \\
\label{H2a} H_{n,ij}^{(ab)}\!\!\!&=&\!\!-\frac{1}{16}J\!\left(1\!-\!\vec
S_i\!\cdot\!\vec S_j\right)\! \left(\!\frac{19}{6}\mp\!
\tau_i^z
\mp\tau_j^z-\!\frac{2}{3}\tau_i^z\tau_j^z\!\right),           \\
\label{H3a} H_{n,ij}^{(ab)}\!\!\!&=&\!\!-\frac{1}{16}Jr_3\!\left(1\!-\!\vec
S_i\!\cdot\!\vec S_j\right)\! \left(\!\frac{5}{2}\mp\!\tau_i^z
\mp\!\tau_j^z+\!2\tau_i^z\tau_j^z\!\right).
\end{eqnarray}
When the spectral weight is evaluated following Eq. (\ref{hefa}),
it is reasonable to try first the MF approximation and to separate
spin and orbital correlations from each other. The spectral
weights require then knowledge of spin correlations along the $c$
axis and within the $ab$ planes (\ref{spins}), as well as the
corresponding intersite correlations
$\langle\vec{\tau}_i\cdot\vec{\tau}_j\rangle$ and
$\langle{\tau}_i^\alpha{\tau}_j^\alpha\rangle$ with
$\alpha=x,y,z$. From the form of the above superexchange
contributions one sees that high-spin excitations
$H^{(\gamma)}_{n,ij}$ support the FM coupling while the low-spin
ones, $H^{(\gamma)}_{2,ij}$ and $H^{(\gamma)}_{3,ij}$, contribute
with AF couplings.

The low-energy optical spectral weight for the polarization along the
$c$ axis $K_{1,exp}^{(c)}$ decreases by a factor close to two when the
temperature increases from $T\simeq 0$ to $T=300$ K \cite{Miy02} ---
this change is much larger than the one observed in LaMnO$_3$ 
\cite{Kov04}. However, the theory based on the MF decoupling of the 
spin and orbital degrees of freedom gives only a much smaller reduction 
of the weight close to 27\%, and has no chance to explain the 
experiment as the maximal possible reduction of $K_1^{(c)}$ found for
$s_c=0$ in the limit of $T\to\infty$ amounts to 33\% \cite{Ole05}.
Note that both spin and orbital intersite correlations change in the
temperature range $0<T<300$ K used in experiment, but this
variation is clearly not sufficient to describe the experimental data.

In contrast, when a cluster method is used to determine the optical 
spectral weight from the high-spin superexchange term (\ref{H1c}) by
including orbital as well as joint spin-and-orbital fluctuations along 
the $c$ axis, the temperature dependence resulting
from the theory follows the experimental data \cite{Kha04}. This may be
considered as a remarkable success of the theory based on the
spin-orbital superexchange model derived for the $R$VO$_3$
perovskites, an the proof that spin-orbital entangled states contribute 
in a cruciat way in the finite temperature regime. 
In addition, the theoretical calculation predicts that the
low energy spectral weight is low along the $c$ axis. The spectral
weight in the $ab$ planes behaves in the opposite way --- it is small
at low energy, and high (but not as high as the low-energy one for the
$c$ axis) at high energy. This weight distribution and its anisotropy
between the $c$ and $ab$ directions reflects the nature of magnetic
correlations, which are FM and AF in these two directions. A more
precise comparison of these theory predictions for the $ab$ polarization
is not possible at present, but we expect that future experiments will 
also confirm them.

\subsection{Peierls dimerization in YVO$_3$}

The third and final example of the spin-orbital entanglement at finite
temperature in the family of vanadate perovskites is the remarkable 
first order magnetic transition at $T_{N2}=77$ K from the $G$-AF to the 
$C$-AF spin order with rather exotic magnetic properties, found in
YVO$_3$ \cite{Ren00}. This magnetic transition is unusual and 
particularly surprising as the staggered moments are approximately
parallel to the $c$ axis in the $G$-AF phase, and reorient above 
$T_{N2}$ to the $ab$ planes in the $C$-AF phase, with some small 
alternating $G$-AF component along the $c$ axis. First, while the 
orientations of spins in $C$-AF and $G$-AF phase are consistent with 
the expected anisotropy due to spin-orbit coupling \cite{Hor03}, the
observed magnetization reversal with the weak FM component remains
puzzling. Second, it was also established by neutron scattering
experiments \cite{Ulr03} that the scale of
magnetic excitations is considerably reduced for the $C$--AF phase
(by a factor close to two) as compared with the exchange constants
deduced from magnons measured in the $G$-AF phase. In
addition, the magnetic order parameter in the $C$-AF phase of
LaVO$_3$ is strongly reduced to $\simeq 1.3\mu_B$, which cannot be
explained by rather small quantum fluctuations in the $C$-AF phase
\cite{Rac02}. Finally, the $C$-AF phase of YVO$_3$ is dimerized.
Until now, only this last feature found a satisfactory explanation
in the theory \cite{Sir03,Sir08}, see below.

We remark that the observed dimerization in the magnon dispersions
may be seen as a signature of {\it entanglement in excited states\/}
which becomes active at finite temperature. The microscopic reason of
the anisotropy in the exchange constants
\begin{equation}
\label{jcdim}
{\cal J}_{c1}\equiv{\cal J}_c(1+\delta_s)\,,\hskip .7cm
{\cal J}_{c2}\equiv{\cal J}_c(1-\delta_s)\,,
\end{equation}
is the tendency of the orbital chain to dimerize, activated by thermal
fluctuations in the FM spin chain \cite{Sir08} which support dimerized
structure in the orbital sector. As a result one finds alternating
stronger and weaker FM bonds along the $c$ axis (\ref{jcdim}) in the 
dimerized $C$-AF phase (with $\delta_s>0$). The observed spin waves may 
be explained by the following effective spin Hamiltonian for this phase
(assuming again that the spin and orbital operators may be disentangled
which is strictly valid only at $T=0$):
\begin{eqnarray}
\label{hcafd}
{\cal H}_{s}&=&{\cal J}_{c}\sum_{\langle i,i+1\rangle\parallel c}
   \left\{1+(-1)^i\delta_s\right\}{\vec S}_{i}\cdot{\vec S}_{i+1}
\nonumber\\
   &+&{\cal J}_{ab}\sum_{\langle ij\rangle\parallel ab}
      {\vec S}_i\cdot{\vec S}_j
   +K_z\sum_i\left(S_i^z\right)^2\,.
\end{eqnarray}
Following the linear spin-wave theory
the magnon dispersion is given by
\begin{equation}
\label{spinw}
\omega_{\pm}({\bf k})\!=\! 2\sqrt{\!\left(2{\cal
J}_{ab}+|{\cal J}_c|+\frac12 K_z \pm {\cal J}_c\eta_{\bf
k}^{1/2}\right)^2 \!\!-\big(2{\cal J}_{ab}\gamma_{\bf
k}\big)^2},
\end{equation}
with
\begin{eqnarray}
\label{gamma}
\gamma_{\bf k}&=&\frac12\left(\cos k_x+\cos k_y\right)\;,  \\
\label{etak}
\eta_{\bf k}&=&\cos^2k_z+\delta_s^2\sin^2k_z\,.
\end{eqnarray}
The single-ion anisotropy term $\propto K_z$ is responsible for the
gap in spin excitations. Two modes measured by neutron scattering
\cite{Ulr03} are well reproduced by $\omega_{\pm}({\bf k})$ obtained
from Eq. (\ref{spinw}) using the experimental exchange interactions:
${\cal J}_{ab}=2.6$ meV, ${\cal J}_c=-3.1$ meV, $\delta_s=0.35$.
We note that a somewhat different
Hamiltonian with more involved interactions was introduced in
ref. \cite{Ulr03}, but the essential features seen in the experiment
are well reproduced already by the present effective spin exchange 
model ${\cal H}_s$, see Eq. (\ref{hcafd}).

The observed dimerization in the magnon spectra in YVO$_3$ motivated
the search for its mechanism within the spin-orbital superexchange
model. Dimerization of AF spin chains coupled to phonons is well
known and occurs in several systems \cite{Joh00}. The spin-Peierls
transition discovered in CuGeO$_3$ \cite{Hase} led to renewed interest
in the dimerization instability of the AF spin chains. In the
spin-orbital model for the $R$VO$_3$ perovskites a similar instability
might also occur without the coupling to the lattice when Hund's
exchange is sufficiently small. In particular, the ground state
at $\eta=0$ may be approximated by the dimerized chain with strong FM
bonds alternating with the AF ones, if such chains are coupled by AF
interactions along the $a$ and $b$ axes \cite{She02} (the 1D chain
would give the entangled disordered ground state as described in Sec.
\ref{sec:enta}).

At realistic values of $\eta>0.10$ the $C$-AF order with FM chains
along the $c$ axis is found in the ground state \cite{Kha01}.
Numerical studies performed at finite temperature have shown that
periodic dimerization of the magnetic exchange exists in a
certain finite temperature range, while the ground state is the
fully polarized and uniform FM state \cite{Sir03,Hor03}. These
findings served as a motivation to investigate the mechanism of
the spin-Peierls dimerization in FM spin chains. The microscopic
1D model which stands for the situation encountered in the $C$-AF
phase of YVO$_3$ reads \cite{Sir08}:
\begin{equation}
\label{SO1}
H_{S\tau}=J\sum_i\left(\vec{S}_i\!\cdot\!\vec{S}_{i+1}+1\right)
\left(\vec{\tau}_i\!\cdot\!\vec{\tau}_{i+1}+\frac{1}{4}-\gamma_H\right),
\end{equation}
where $\gamma_H$ is stands for the contribution due to the high-spin
states proportional to the Hund's exchange (\ref{eta}) and stabilizes 
FM spin order. While the spin and orbital operators are disentangled 
in the FM ground state, one may consider a coupled FM spin chain to an
orbital chain with interactions which favor the AO order, as realized
in the $C$-AF phase. The exchange interactions along the spin (orbital)
chain depend on the orbital (spin) correlations, and their modulation
may be described by $\delta_{s}$ and $\delta_{\tau}$ parameters 
(\ref{jcdim}). They can be found from a self-consistent solution of the 
coupled MF equations for spin and orbital correlations, and one finds 
indeed dimerized spin and orbital chains in a finite range of 
temperature \cite{Sir08}.

Summarizing, spin-orbital entanglement in the excited states is also
responsible for the exotic magnetic properties of the $C$-AF phase
of YVO$_3$. They arise from the coupling between the spin and
orbital operators which triggers the dimerization of the FM
interactions as a manifestation of a universal instability of FM
chains at finite $T$, which occurs either by the coupling to the
lattice or to purely electronic degrees of freedom \cite{Sir08}.
This latter mechanism could play a role in many transition metal
oxides with (nearly) degenerate orbital states.

\section{Coexisting charge and orbital order}
\label{sec:co}

The first step towards understanding the doped systems with
orbital degrees of freedom is the question concerning possible QP
states deciding about coherent hole propagation in the orbitally
ordered background. As discussed in Sec. \ref{sec:cup}, a single
hole doped into the AF background as in CuO$_2$ planes of
La$_2$CuO$_4$ may propagate through the lattice because it couples
to quantum spin fluctuations and becomes dressed with a "cloud" of
magnons \cite{Bul68}. This results in the new energy scale
$\propto J$ in place of the hole hopping $t$, so the hopping is
strongly renormalized. The QP which forms after the hole is doped
in the AF background is called a {\it spin polaron} \cite{Mar91}.
A more complex situation can occur in the systems with partly
filled degenerate orbitals, where a doped hole may not only couple
to magnons but also couples to crystal-field excitations \cite{Zaa93}. 
In addition, QP states with higher spin states may occur, as for
instance a triplet QP in case of an $S=1/2$ antiferromagnet doped 
by a single electron in the orbitally degenerate background 
\cite{Zaa92}. This motivates two questions in the theory: 
($i$) whether orbital excitations could couple as well to the moving 
hole and generate a new energy scale, as the magnons do, and 
($ii$) whether spin-orbital entanglement has any important 
consequences for the hole dynamics. Both of them were addressed in the 
orbital $t$--$J$ model for $e_g$ electrons \cite{Dag07}, and in the 
analogous models for $t_{2g}$ orbitals developed recently, see below.

Two situations with a hole doped into an AO ordered background
were considered in the past: 
($i$) a hole doped into an $ab$ plane of LaMnO$_3$ \cite{vdB00} which 
has an AO order of $e_g$ orbitals in the ground state, and 
($ii$) a hole doped into an $ab$ plane with an AO order of $t_{2g}$ 
orbitals and FM spin order \cite{Dag08}, as realized for instance in 
Sr$_2$VO$_4$. 
In the first case it was shown that the orbitons have in general 
a gap and have a lower dispersion than the magnons. Therefore, the
quantum effects are weak but but a hole can move by interorbital
hopping processes. While the constraint of creating no double
occupancies has to be obeyed along the hole hopping, the bandwidth
is strongly renormalized with respect to that suggested by the
LDA+$U$ approach \cite{vdB00}. Such interorbital hopping processes
are absent in the $t_{2g}$ ordered background with alternating
$yz/zx$ orbitals in an $ab$ plane, and due to the specific
$t_{2g}$ orbital symmetries the orbitons are dispersionless. Thus
the string picture \cite{Wro08} dominates the character of the
$t_{2g}$ orbital polarons even more than in the case of systems
with $e_g$ orbital degrees of freedom.

An intriguing question in this context addressed only recently is
whether spin quantum fluctuations can still contribute to the QP
spectral properties when both types of order, spin and orbital,
alternate in an $ab$ plane, as for instance in the $C$-AF phase of
$R$VO$_3$ perovskites. A crucial observation for the spectral
properties of a hole doped into the entangled AF/AO background is
a simultaneous excitation of a magnon and an orbiton when a hole
moves by a single step in the lattice \cite{Woh09}. This dominates
the behavior of the hole doped in such an entangled state, because
the orbitals confine the hole motion by forcing the hole to
retrace its path which implies that the hole motion by its
coupling to the quantum spin fluctuations is prohibited. Thus,
{\it the string-like potential which acts on the hole is induced
by the orbitals\/} although it has a joint spin-orbital character.
Hence, this important feature of the orbitally induced string
formation could be understood as a topological effect. This
happens even if the energy of the orbital excitations is turned to
zero, i.e., when the hole moves in the orbital sector
incoherently. Hence, the mere presence of orbitals is sufficient
to obtain the (almost) classical behavior of a hole doped into the
ground state with AF/AO order. This result, in connection with the 
fact that the mother-compound of the superconducting iron-pnictides
shows a variety of spin-orbital phenomena \cite{Kru09}, suggests
that further investigation of the hole propagation in spin-orbital
systems is a fascinating subject for future studies.

The properties of doped $R_{1-x}$Sr$_x$VO$_3$ systems are puzzling
and it is not understood until now why ($i$) La$_{1-x}$Sr$_x$VO$_3$ is
insulating in a broad range of doping below $x_c=0.18$, and ($ii$) why 
the AF order survives even for $x>x_c$ when the system becomes
metallic and looses the AO order in the $ab$ planes \cite{Fuj06}.
The gradual changes of the optical conductivity under increasing
doping demonstrate that the anisotropy between the $ab$ and $c$
direction decreases, but surprisingly is not completely lost even in 
the metallic regime. The differences observed in the optical
conductivity and raman scattering spectra between
La$_{1-x}$Sr$_x$VO$_3$ and Y$_{1-x}$Sr$_x$VO$_3$ suggest that the
orthorhombic lattice distortion plays also here a very important
role and influences the hole dynamics \cite{Fuj08}. It has been
argued that the $C$-AF phase is more robust under hole doping
\cite{Ish05} and survives in a broad doping range \cite{Woh05},
but a complete understanding of doped vanadate perovskites awaits
a more careful theoretical study.

Doped $R_{1-x}$(Sr,Ca)$_x$MnO$_3$ systems are studied much longer
and they are better understood. The FM metallic state is induced
by doping via the double exchange mechanism \cite{Dag01} which was
also formulated for degenerate $e_g$ orbitals \cite{Kho99}, and
the phase diagrams of the doped perovskite systems show a remarkable 
sequence of magnetic phases \cite{Che95}, from the $A$-AF phase, 
through the insulating and metallic FM phase, towards the $C$-AF 
phase and $G$-AF phase in the highly doped regime. Similar (but not 
the same) sequence of magnetic phases was reported in the bilayer
La$_{2-2x}$Sr$_{1+2x}$Mn$_2$O$_7$ systems \cite{Lin00}. As in the 
$R_{1-x}$Sr$_x$VO$_3$ perovskites, also in the layered systems the
orbital ordered (or liquid) state determines whether the intersite
spin correlations are AF or FM, as shown for the monolayer \cite{Dag04} 
and for the bilayer \cite{Dag06}using a $t$--$J$-like
model which includes orbital degeneracy. These model calculations
illustrate as well the complementarity of spin and orbital order
expressed by the Goodenough-Kanamori rules \cite{Kan59}. In these
systems the short-range charge order gradually develops with
increasing doping in the realistic parameter regime \cite{Ros07}.
However, more complete models including the charge transfer
physics are necessary to describe the features observed in the
optical spectra, as for instance in insulating LaSrMnO$_4$
\cite{Gos08}.

Although there is no complete understanding of the phase diagram
and in particular of the mechanism of the metal-insulator
transition which leads to the colossal magnetoresistance until now, 
a lot of progress could be made using model Hamiltonians. It was 
recognized that the orbital degeneracy plays a crucial role both in 
the double exchange \cite{Kho99} and for the interactions with the
lattice due to the JT effect \cite{Sti07}, and phase diagrams
which resemble the qualitative behavior of the doped manganites
were obtained \cite{Sti08}. However, a more realistic treatment
requires also electron correlations among $e_g$ electrons which
are more difficult to implement \cite{Feh04}. In contrast to the
nondegenerate Hubbard model, the orbital Hubbard model for FM
manganites does not show an instability towards the orbitally
polarized FO state and one finds instead the disordered orbital 
liquid ground state \cite{Fei05}. This concept was crucial in 
explaining the doping dependence of the stiffness constant in the 
FM La$_{1-2x}$Sr$_{x}$MnO$_3$ manganites \cite{Per96}, but for a
quantitative explanation both the double exchange due to
correlated $e_g$ electrons and the superexchange due to $t_{2g}$
core spins had to be included \cite{Ole02}. This approach had also
a remarkable success \cite{Ole03} in explaining the observed
magnon dispersion and the doping dependence of the magnetic
exchange constants in the La$_{2-2x}$Sr$_{1+2x}$Mn$_2$O$_7$
systems, including the observed phase transition from the FM to
the $A$-AF structure \cite{Per99}.

As expected, the orbital order melts in general when the
manganites are doped, there are cases when a different
type of orbital order coexisting with charge order emerges again
at half doping. The famous case is the so-called (charge exchange)
CE phase in half-doped ($x=0.5$) manganites \cite{Goo55}, where
the two-sublattice charge order coexists with orbital order on the
sites with the majority of $e_g$ electron charge, and the FM
zig-zag chains staggered in $ab$ planes. Although the double
exchange provides some arguments justifying the stability of this
complex type of order \cite{Kha99} which competes with the FM
phase in the relevant parameter regime and wins for sufficiently
large and realistic AF superexchange between the $S=3/2$ $t_{2g}$
core spins \cite{Bre07}, the problem is subtle and the range of
parameters with the CE phase in the ground state is rather narrow.
In particular, this phase is destabilized by intersite Coulomb
interaction \cite{Dag06}, and the JT distortions play an important
role in stabilyzing it \cite{Ali03}. We emphasize again that 
AF interactions between $t_{2g}$ electrons are small --- for this 
case the CE phase was found in the charge-ordered phase using a 
finite-temperature diagonalization technique \cite{Bal04}. The
mechanism invoked there to stabilize the CE phase is subtle and 
employes the cooperative JT interaction between next-nearest Mn$^{3+}$ 
neighbors mediated by the breathing mode distortion of Mn$^{4+}$
octahedra and displacements of Mn$^{4+}$ ions. It is worth noting 
that the topological phase factor in the Mn-–Mn hopping \cite{Hot00}
leading to gap formation in 1D models \cite{Kha99} for the CE phase,
as well as the nearest neighbor JT coupling, are not able to produce 
the observed zigzag FM chains for the realsitic parameters  
\cite{Bal04}. Recent x-ray structural analysis of 
Pr$_{0.5}$Ca$_{0.5}$MnO$_3$ and Eu$_{0.5}$Ca$_{1.5}$MnO$_4$ suggest
that the orbital shape and the charge disproportionation are sensitive 
to the dimension of Mn--O network \cite{Oku09}, which together with 
the possible different role of the JT effect in both compounds poses
new interesting questions in the theory.

The controversy about the
nature of the charge order in this phase which arose due to
signatures of Zener polarons observed in the neutron data
\cite{Ala00} seems to be resolved now in favor of the more
conventional picture of zig-zag chains \cite{Sen06}. Melting of
this composite order with increasing temperature is fascinating
and the magnetic order disappears first, giving coexisting charge
and orbital order in the intermediate temperature regime, before
both melt resulting in a disordered phase \cite{Sen08}. This
suggests that the link between (weak) charge and orbital order is
particularly strong here, similar as in the magnetite below the
Verwey transition \cite{Pie06}. Recently, charge ordered AF phase
was also reported in La$_{1.5}$Sr$_{0.5}$CoO$_3$, and preliminary
theoretical concepts in the framework of spin-orbital physics were
also presented \cite{Hel09}.

Stripe phases appear also in the doped systems with active orbital
degrees of freedom, but are qualitatively different from the ones
observed in the cuprates, see Sec. \ref{sec:cup}. Stripe order was
found in doped manganites \cite{Che98} and also discovered in doped 
La$_{2-x}$Sr$_x$NiO$_4$ nickelates about the same time as in the 
cuprates \cite{Tra95}. However, in contrast to the cuprates the 
stripes in La$_{5/3}$Sr$_{1/3}$NiO$_4$ are diagonal and contain one 
(and not half) hole per unit cell\cite{Wak09}. Intriguing features
seen in the spin excitation spectra of La$_{2}$NiO$_{4+\delta}$
nickelates were reported recently \cite{Fre09} which suggest that
the inward dispersion, seen also in cuprates, has a common origin
in stripe phases. Simulations performed within
the LDA+$U$ approach suggest that a subtle interplay 
between the charge and spin order and octahedral distortions is
essential for the formation of an insulating state \cite{Sch09}.
Preliminary HF calculations emphasize the importance of orbital
degeneracy in the case of nickelates for the realistic $e_g$
hopping model \cite{Rac06a}, where one finds indeed that diagonal
stripes with the observed filling (of one hole per unit cell) are
more stable than other phases, in contrast to the predictions of
the degenerate Hubbard model with diagonal hopping (which does not
agree with experiment). It remains a challenge for the theory to
develop a more complete theory of the stripe phases in the
nickelates, including the electron correlations and the coupling
of $e_g$ electrons to the lattice distortions, and to understand 
better differences between the stripe phase in the nickelates and
in the cuprates .

\section{Summary and open problems}
\label{sec:summa}

Summarizing, charge order is common in doped transition metal
oxides and arises even in absence of intersite Coulomb
interactions while the kinetic energy of doped charges (holes or
electron) competes with the magnetic superexchange. Under such
circumstances stripe phases realized in the cuprates, nickelates,
and manganites are favored as then the two above energies are
optimized simultaneously in the domain walls and in the magnetic
domains between them. When also orbital degrees of freedom are
active, the charge order is accompanied by certain, usually weak,
orbital order. Good examples of this composite type of
charge-and-orbital order are the low temperature phase of 
Fe$_3$O$_4$, and the CE phase in the half-doped manganites.

In the perovskite lattice the orbital order is stabilized easier
in the correlated insulators with orbital $e_g$ degrees of freedom, 
as then the superexchange and the interaction with the lattice act 
supporting each other \cite{Fei99}, than in the ones with the $t_{2g}$ 
active orbitals, where the interactions with the lattice in general 
compete with the superexchange \cite{Hor08}. As a result, the orbital 
and magnetic transition occur independently from each other and at 
quite different temperatures in the $e_g$ systems, while the case of 
the $R$VO$_3$ perovskites is an example of the proximity and interplay 
of the magnetic and orbital phase transition. Both these different 
situations were successfully described within spin-orbital 
superexchange models with added interactions with the lattice.

A qualitative difference between the orbital order in $e_g$ and
$t_{2g}$ systems is that the orbital order is more robust in $e_g$
case and may be usually treated by classical (MF) approaches,
while $t_{2g}$ orbitals may easier fluctuate and thus couple also
easier to the spin degrees of freedom. It is for this reason that
composite spin-orbital fluctuations occur in correlated titanates
and vanadates insulators. Although such fluctuations are quenched
in the ground state of these systems for realistic parameters,
they develop at increasing temperature due to the presence of
excited stated with spin-orbital entanglement. For instance, such
composite spin-orbital fluctuations are responsible for the
temperature dependence of the optical spectral weights in LaVO$_3$
\cite{Kha04} and trigger spin-orbital dimerization in the $C$-AF
phase of YVO$_3$ in the intermediate temperature regime
\cite{Hor03}. Similarly interesting orbital ordered states are
also found in the perovskite ruthenates --- as an example we mention
here the puzzling low temperature electronic and structural behavior 
recently discovered in PbRuO$_3$ \cite{Kim09}.

The microscopic mechanism of melting of the orbital order in doped
systems is a very challenging problem in the theory and could not
be understood until now. The main difficulties follow from
disorder and the necessity of using Monte Carlo techniques. In
this way it could be concluded that a strong competition between
the FM metallic and the AF charge-ordered insulating states takes
place \cite{Yu08}. This competition influences the transport
properties and leads to short-range spin and charge correlations
which evolve with time. It is intriguing to what extent this
complex situation in the metallic phase influences the magnetic
excitations of the system. The magnetic excitations in the
metallic FM phase of several doped manganites soften at the zone
boundary and may be described by the Heisenberg model with the
nearest neighbor $J_1$ and fourth-nearest neighbor $J_4$ exchange
coupling \cite{End05}. Its microscopic origin is controversial and
two different concepts were proposed to explain the experimental
observations: ($i$) quantum fluctuations of the planar $x^2-y^2$
orbitals associated with the $A$-AF phase \cite{Kil00}, and ($ii$)
the $(3z^2-r^2)$-type orbital fluctuations \cite{End05}. It was
also shown that the ratio of $J_4/J_1$ changes along the
$A_{1-x}A'_x$MnO$_3$ manganites (with $A$ and $A'$ being the
rare-earth and alkaline rare-earth ions), while the stiffness
constant is almost universal and has only very weak dependence on
the chemical composition for a fixed doping $x$ \cite{Ye06}. Both
theoretical models have difficulties to explain the experimental
data --- an incorrect dispersion along the (111) direction follows
from the first one \cite{Kil00}, while the second one predicts a
spectacular doping dependence of $J_4/J_1$ which is not observed.
Thus, in spite of its remarkable success in the simplest situation
\cite{Ole02}, the complete theoretical explanation of the magnon 
dispersion in the metallic FM manganites within the orbital liquid 
state remains a challenging problem in the theory. Even more puzzling
are the magnons in the insulating FM phase, where several branches
with a staircase-like spectrum have been observed \cite{Mou09}.

Recent studies of the phase diagram of the $R$MnO$_3$ manganites
include the effect of orthorhombic distortions within theoretical
models with spin superexchange and the Dzyaloshinsky-Moriya
interaction responsible for the multiferroic behavior
\cite{Moc09}. The usual approach so far is to develop an effective
spin model including the spin-lattice coupling, leading to the
electric polarization \cite{Mos06}. It seems that an explicit
treatment of the orbital degrees of freedom could provide a better
understanding of the observed phenomena.

Another active direction of research in the field of transition
metal oxides is the search for novel quantum phenomena, including
more examples of quantum spin-orbital entanglement. They could be
found in frustrated lattices, and here we mention briefly only the
triangular lattice. An interesting case and good candidate for a
spin-orbital liquid might be LiNiO$_2$ with a triangular lattice
of Ni$^{3+}$ ions ($d^7$ configuration with $S=1/2$ spins) and no
magnetic or orbital order down to very low temperatures. First, it
was argued that a model based on symmetry arguments \cite{Ver04},
characterized by a large number of low-lying singlets associated
to dimer coverings of the triangular lattice, could explain the
absence of any type of ordered phase in LiNiO$_2$. Second, it was
shown that excited states on oxygens along the 90$^{\circ}$ bonds
are crucial in the superexchange and they change the balance
between different terms in the Hamiltonian, making the orbital
interactions stronger than the spin ones \cite{Rei05}. In any
case, interplane JT coupling seems to be too weak in LiNiO$_2$ to
stabilize the orbital long-range order, and the microscopic reason
of disorder could be alone due to strongly frustrated orbital
interactions on the triangular lattice in (111) planes
\cite{Rei05}, which resemble the compass model. A completely
different situation is encountered in the $d^1$ spin-orbital model
with active three $t_{2g}$ orbitals on the triangular lattice, 
as realized in NaTiO$_2$, where a
spin-orbital disordered liquid state is more likely \cite{Nor08}.

As a final remark, we would like to mention recent experimental
studies of Ni-based superlattices \cite{Cha06}. They stimulated
progress in the theory which predicts that, in analogy to the gain
of kinetic energy in the layered manganites \cite{Mac99}, the
correlated $e_g$ electrons in the NiO$_2$ planes develop a planar
$(x^2-y^2)$-like orbital order in LaNiO$_3$/La$M$O$_3$
superlattices (with $M$=Al, Gd, Ti) \cite{Cha08}. It may be
expected that future studies of the systems of reduced
dimensionality will provide more unexpected properties in the near
future, and could lead to developing functional materials, using
both charge and orbital degrees of freedom.

\acknowledgments

It is a great pleasure to thank all my collaborators for numerous
insightful discussions which contributed to my present understanding
of the subject --- I thank in particular my friends, Lou-Fe' Feiner,
Peter Horsch, Giniyat Khaliullin, Jan Zaanen, M. Daghofer, B. Normand, as well 
as the members of the research group at the Jagellonian University:
J. Ba\l{}a, W. Brzezicki, M.~Raczkowski, K.~Ro\'sciszewski, 
and K.~Wohlfeld.
We acknowledge support by the Foundation for Polish Science (FNP),
and by the Polish Ministry of Science and Higher Education under
Project No.~N202 068 32/1481.

\end{document}